# Conductance switching at the nanoscale of diarylethene derivatives self-assembled monolayers on $La_{0.7}Sr_{0.3}MnO_3$

L. Thomas,[1] D. Guérin,[1] B. Quinard,[2] E. Jacquet,[2] R. Mattana,[2] P. Seneor,[2]
D. Vuillaume,[1,*] T. Mélin[1] and S. Lenfant.[1,*]

*1. Institute for Electronics Microelectronics and Nanotechnology (IEMN), CNRS, Univ. Lille, 59652 Villeneuve d'Ascq, France.*

*2. Unité Mixte de Physique, CNRS, Thales, University Paris Sud, Université Paris Saclay, 91767 Palaiseau, France.*

* Corresponding authors : stephane.lenfant@iemn.fr; dominique.vuillaume@iemn.fr

## Abstract.

We report on the phosphonic acid route for the grafting of functional molecules, optical switch (dithienylethene diphosphonic acid, DDA), on $La_{0.7}Sr_{0.3}MnO_3$ (LSMO). Compact self-assembled monolayers (SAMs) of DDA are formed on LSMO as studied by topographic atomic force microscopy (AFM), ellipsometry, water contact angle and X-ray photoemission spectroscopy (XPS). The conducting AFM measurements show that the electrical conductance of LSMO/ DDA is about 3 decades below that of the bare LSMO substrate. Moreover, the presence of the DDA SAM suppresses the known conductance switching of the LSMO substrate that is induced by mechanical and/or bias constraints during C-AFM measurements. A partial light-induced conductance switching between the open and closed forms of the DDA is observed for the LSMO/DDA/C-AFM tip molecular junctions (closed/open conductance ratio of about 8). We show that, in the case of long-time exposition to UV light, this feature can be masked by a non-reversible decrease (a factor of about 15) of the conductance of the LSMO electrode.

## Introduction.

In organic spintronics, $La_{0.7}Sr_{0.3}MnO_3$ (LSMO) is a widely used substrate since it is an air-stable ferromagnetic oxide with a high spin polarization. Several organic semiconductors and polymers have been deposited on LSMO and spin-polarized electron transfer was demonstrated (see a review in 1, and references therein). Recently, molecular tunnel junctions were successfully fabricated by the chemical grafting of alkylphosphonic derivatives self-assembled monolayers (SAM) on LSMO.[2-4] Tunnel magnetoresistance (TMR) from few tens to $10^4$ % was reported, with a significant stability up to high voltages (few volts on nm-thick SAM).[5, 6] The alkyl chains that were used were simply insulating tunnel barriers. Now, functional molecules are required to move towards more elaborated molecular devices,[7, 8] e.g., redox molecules for memory, photochromes for electro-optical molecular devices. In this latter case, diarylethene (DAE) is acknowledged as an archetype of optically driven molecular switches,[9-12] for which the closed form (DAE-c) is more electrically conducting than the open form (DAE-o). As such, for DAE molecular junctions on gold electrodes, conductance switching ratios from 3-4 up to about 100 between the closed and open forms were theoretically predicted,[13] and measured.[14]

In parallel, as in many oxide materials of interest for memory and memristor applications,[15, 16] "memory" effects (electrical conductance switching upon bias stress) were also reported for LSMO films.[17-19] Thus, these features make possible to combine LSMO and DAE for the design and study of multifunctional optical-spintronic devices (i.e., responsive to several, electric, magnetic and optical stimuli).

Here, we report on the successful formation of SAMs of dithienylethene diphosphonic acid (DDA) on LSMO substrates. The physicochemical characterizations (ellipsometry, contact angle, X-ray photoelectron spectroscopy, atomic force microscopy) of the SAMs reveal the formation of compact, defect-free, stoichiometric SAMs of DDAs on LSMO. The electronic transport properties at the nanoscale were investigated by ambient and ultra-high vacuum (UHV) conducting atomic force microscopy (C-AFM) separating the contribution from the LSMO electrodes and the DDA SAMs. In particular, we demonstrate that the presence of the SAMs suppresses the conductance switching of the LSMO substrates upon bias/force constraints during C-AFM measurements. We show the optically induced conductance switching of the DDA SAMs on LSMO between the open and closed forms with a moderate conductance ratio (about 8). Moreover, we observe that, under certain conditions (e.g., long time ultraviolet illumination), this DDA switching can be masked by a larger optical switching of LSMO substrate conductance (conductance ratio of about 15).

## Device fabrication and characterization methods.

The LSMO thin film electrodes (20 nm thick) were fabricated by pulsed laser deposition on $SrTiO_3$ substrates (area ≈ 1 $cm^2$) according to a process already reported elsewhere.[3] Prior to the formation of the DDA SAM, the LSMO surface was ultrasonically cleaned 5 min in deionized (DI) water, followed by 5 min in ethanol and dried under a $N_2$ stream. The DDA (molecular structure in scheme 1) was synthesized following the method of Reisinger et al.[20] with slight modifications described in the supporting information. The DDA was designed to operate as a standard diarylethene derivative switch, only the end groups have been changed to phosphonic groups. The closing/opening of the diarylethene core enhance/disrupt the electronic conjugation across the molecule and modulating the conductance, with a higher conductance in the closed form.[10] This compound was successfully characterized by NMR spectroscopy as well as by mass spectrometry (see supporting information). Moreover, we checked by UV-vis spectroscopy that DDA molecules could switch reversibly in solution by irradiation at 365 nm and 470 nm (Fig S1, supporting information). SAMs were prepared by immersing a LSMO substrate in a millimolar solution of DDA in ethanol for 24 h in the dark (see detail in the supporting information).

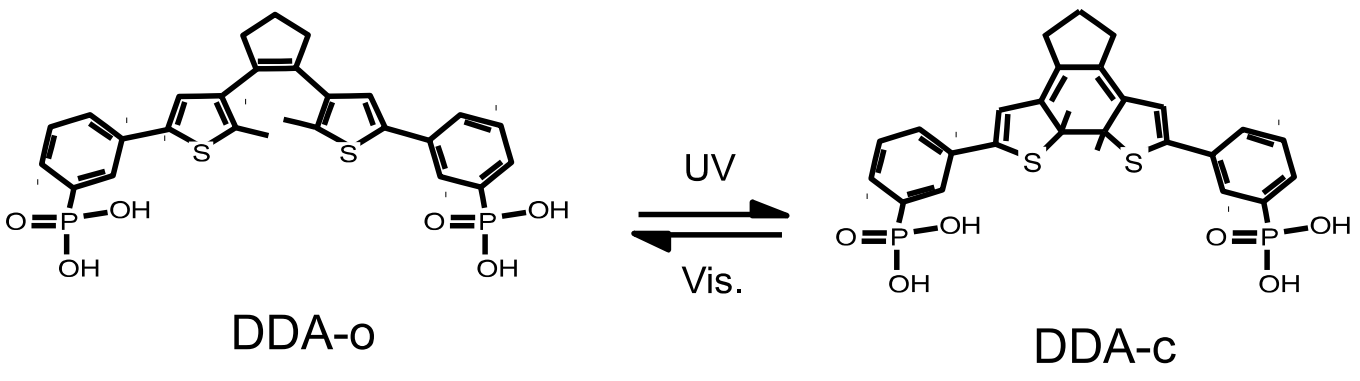

**Scheme 1.** Open and closed forms of the DDA molecules.

SAMs were characterized by ellipsometry and deionized (DI) water contact angle (see supporting information). X-ray photoemission spectroscopy (XPS) spectra of the bare LSMO and LSMO/DDA structures were measured to assess the presence of the DDA molecules and possible modifications of the LSMO surfaces upon SAM deposition (see supporting information).

The electron transport properties at the nanoscale were measured by C-AFM in air and (or under a dry $N_2$ flux) and in ultra-high vacuum (UHV). The surface topography of the LSMO and LSMO/DDA samples were measured by AFM (see supporting information).

To trigger the switching of the DDA molecules from the open (DDA-o) to the closed (DDA-c) forms, we irradiated the samples with Thorlabs LEDs coupled with an optical fiber in UV light (at 365 nm) and blue light (at 470 nm), or let in the dark for a long period (12h) for the DDA-c to DDA-o switching (see supporting information).

## Results and discussion.

### *Structure of the DDA SAM and DDA/LSMO interface.*

The thicknesses of the SAMs measured by ellipsometry on several samples are between 1.7 and 2.0 (±0.2) nm. The water contact angles of the bare LSMO substrates are in the 27 to 37° (±2°) range, and they increase to 45 - 68° (±2°) after the SAM deposition. Given the length of the DDA molecule (1.9 nm determined by geometry optimization, MOPAC software[21] with PM3 parametrization), these values are consistent with a DDA molecule grafted by a single phosphonic end and with an average tilt angle to the surface normal between 0 and 47°. The water contact angles measured on the DDA SAMs are in agreement with SAMs exposing both a phosphonic and phenyl groups at their surfaces[22] (the technique is sensitive to chemical species down to few Å inside the SAM[23]). These results indicate the formation of a compact SAM. Figure 1 shows the topographic AFM images of the bare LSMO (Figs. 1a-b) and LSMO/DDA samples (Figs. 1c-d) which are similar. Both samples have a root-mean-square (rms) roughness of 0.6-0.8 nm, with a maximum height variation of about 2 nm (Figs. 1b and 1d) and the same typical topographic features (e.g., bumps and holes with a typical size in the range 20-100 nm, Figs. 1b and 1d). This result indicates the formation of rather conformable, uniform, SAM without large defects (i.e., uncovered zones or multilayers/aggregates). The rms roughness for the LSMO and LSMO/DDA samples are similar to the values reported for LSMO made with the same process and then coated with SAMs of aklylphosphonic molecules.[2, 6]

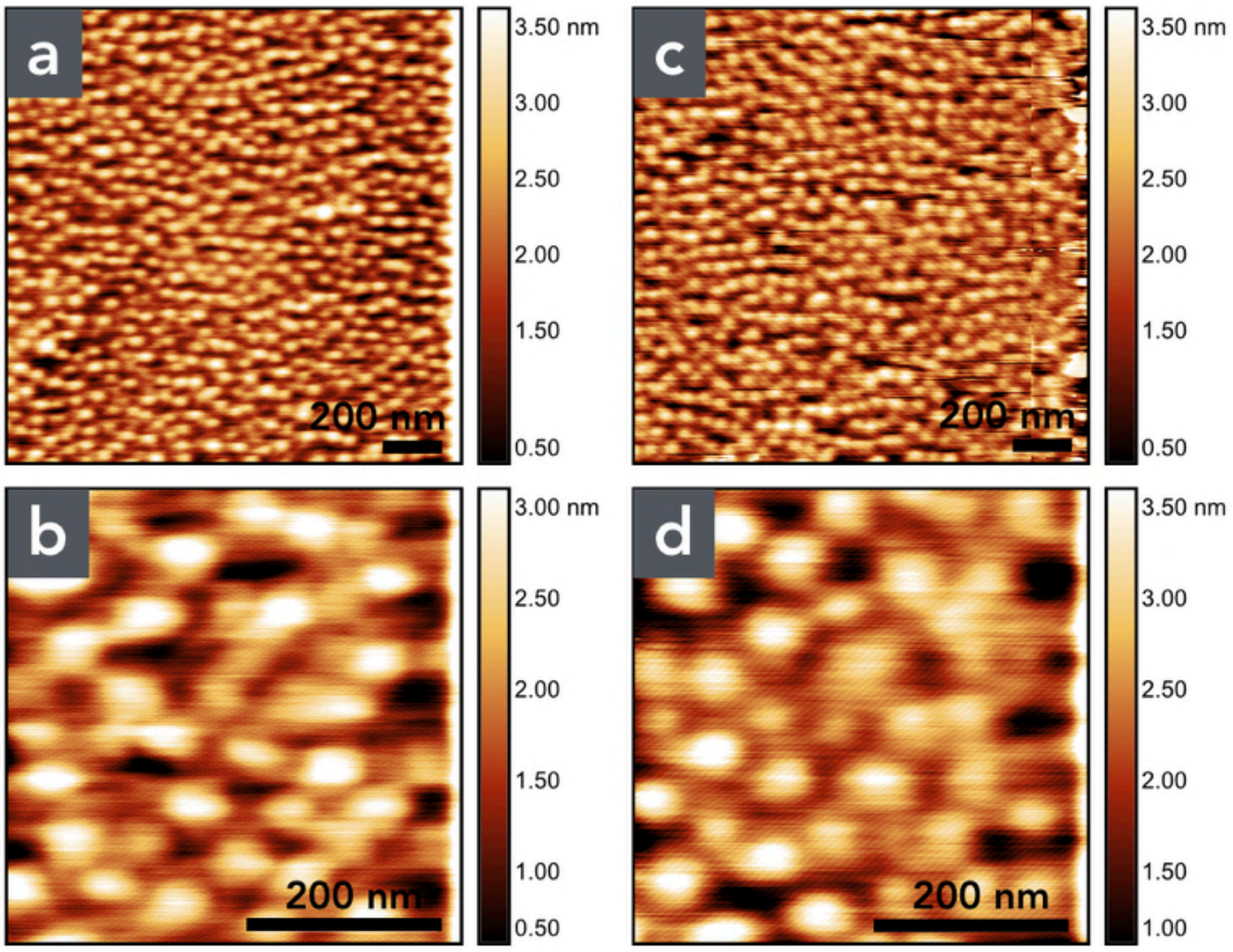

***Figure 1.*** *Topographic AFM images (contact mode, loading force 30 nN) of the LSMO and LSMO/DDA samples :* ***(a)*** *LSMO (2 x 2 μm²),* ***(b)*** *LSMO (0.5 x 0.5 μm²),* ***(c)*** *LSMO/DDA (2 x 2 μm²),* ***(d)*** *LSMO/DDA (0.5 x 0.5 μm²).*

The XPS measurement of the LSMO surface shows the La3d, Sr3d, Mn2p and O1s peaks, with a residual C1s carbon contamination (see Fig. S2 in the supporting information). The peaks were deconvoluted in several contributions and fitted by Voigt or Gaussian functions to determine the stoichiometry of the samples from the corresponding peak areas (see details in the supporting information). We obtain a stoichiometry $La_{0.7}Sr_{0.33}Mn_{0.85}O_{2.82}$ with a slightly rich Sr surface (+10%) in agreement with ref. 24 and a slight Mn deficiency (-15%) - see Table S1 in the supporting information. After the DDA molecules grafting, a similar analysis of the XPS spectra (Fig. S3 in the supporting information) gives a stoichiometry $La_{0.7}Sr_{0.32}Mn_{0.71}O_{3-\delta}$ (here, $\delta$ cannot be determined due to the energy overlapping of O1s in the LSMO and in the DDA molecules) - Table S2 in the supporting information. Since the magnetic and electronic properties of LSMO depend on the Mn oxidation states, we focused on a more detailed analysis of this element.

For the bare LSMO sample, the $Mn2p_{3/2}$ peak is decomposed into two peaks at 641.3 eV and 643 eV corresponding to the oxidation states $Mn^{3+}$ and $Mn^{4+}$, and respectively 652.8 and 655 eV, for the $Mn2P_{1/2}$ peak.[25-27] The ratio of the peak amplitude $Mn^{3+}/Mn^{4+}$ is 0.92. For the LSMO/DDA sample, we have a ratio $Mn^{3+}/Mn^{4+}$ =1.22, thus the DDA grafting leads to a 32% increase of the $Mn^{3+}/Mn^{4+}$ ratio. Such an increase was also previously observed for the grafting of alkylphosphonic acid molecules on LSMO,[3] but without hindering spin dependent injection and transport through the LSMO/alkylphosphonic/Co molecular junctions for which tunnel magnetoresistance was clearly measured.[5, 6]

We clearly observed the P2s (190.6 eV), the $S2p_{3/2}$ (164 eV) and $S2p_{1/2}$ (165.3 eV) peaks of the DDA molecules (Fig. S3 in the supporting information). We also note a strong increase of the C1s peak at 284.8 eV. From the calculated area below the peaks, we determine the C/S and P/S ratios of 15.9 and 1, respectively (theoretical values from the composition of the DDA molecule : 13.5 and 1) - see Table S2 in the supporting information. Thus, we conclude that the DDA molecules are present on the LSMO surface, the slightly higher C/S ratio being due to carbon contamination before or during the grafting process.

In summary, ellipsometry, AFM, water contact angle and XPS measurements show that we have formed compact SAMs of DDA molecules, with minor changes of the LSMO surface stoichiometry, preserving the possibility to study spin-polarized electron transport through these devices based on more complex or functional molecules than simple alkyl chains as previously reported.[2, 3, 5, 6]

***Electron transport properties at the nanoscale.***

The electron transport properties of the LSMO/DDA/metal junctions were measured by C-AFM (metal tip = PtIr or Pt) in UHV and compared to the bare LSMO substrate (Fig. 2). A large number (around one thousand) of current-voltage traces were recorded at different locations on the samples (see the supporting information for details) to construct 2D histograms (I-V curves, Figs.

2a and 2d for the LSMO and LSMO/DDA samples) and 1D current histograms at a given bias. The current histograms (Figs. 2b, 2c for LSMO and 2e and 2f for LSMO/DDA) are fitted by a log-normal distribution with a log-mean current, log μ, and log-standard deviation, log σ. As expected, the grafting of the SAM of DDA molecules induces a large decrease of the current by about 3 decades from log μ = -9.13 (i.e. $7.4 \times 10^{-10}$ A) at 0.5 V and log μ = -8.83 ($1.5 \times 10^{-9}$ A) at -0.5V for the LSMO/PtIr junctions to log μ = -11.98 ($1.05 \times 10^{-12}$ A) and log μ = -11.86 ($1.4 \times 10^{-12}$ A) at 0.5 and -0.5 V, respectively, for the LSMO/DDA/PtIr junctions. In all cases, we observe a similar and large standard deviation (about two decades of current at the FWHM of the distributions), which can be attributed to the spatially inhomogeneous conductivity of the LSMO substrates (see Fig. S4 in the supporting information). Note that we can also decompose these large distributions in two peaks (see later) possibly ascribed to the presence of the two forms of the DDA (open and closed) in the SAM.

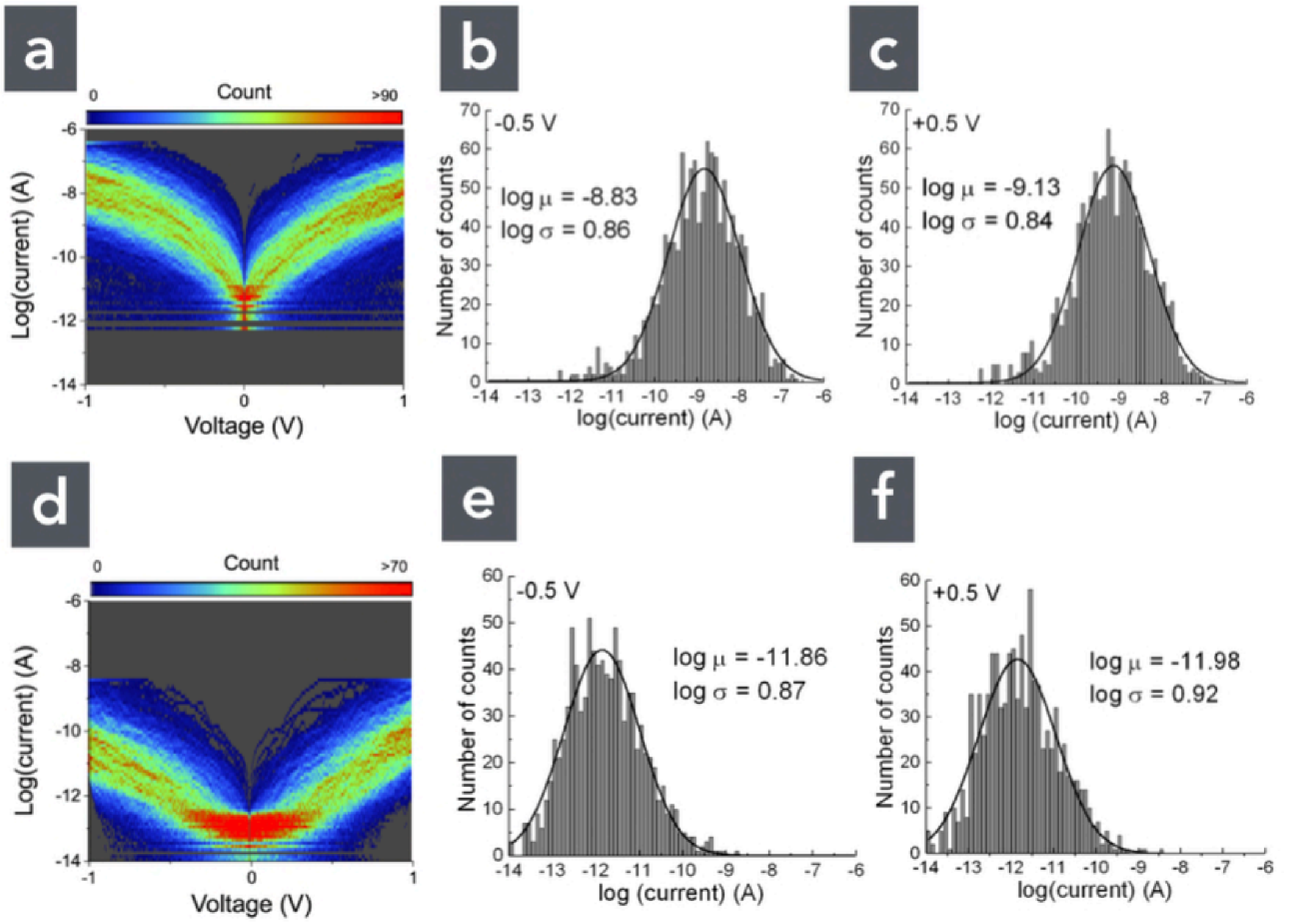

***Figure 2.*** ***(a)*** *Current voltage (I-V) 2D histogram for the LSMO samples, 1200 I-V traces, and* ***(b,c)*** *corresponding 1D current histograms at -0.5 V and 0.5 V.* ***(d)*** *Current voltage (I-V) 2D histogram for the LSMO/DDA samples, 974 I-V traces, and* ***(e,f)*** *corresponding 1D current histograms at -0.5 V and 0.5 V. All measurements by C-AFM in UHV at a loading force of 30 nN. The black lines are the fits with a log-normal distribution. The fit parameters, log-mean current (log μ) and log-standard deviation (log σ) are given in the figures.*

Before testing the conductance switching behavior of the DDA molecules grafted on the LSMO substrates, we studied the conductance switching of the LSMO substrates. We used C-AFM in air and UHV (since oxygen is known to play a key role in this switching behavior)[17-19] and we recorded successive current images by enlarging the scanned area for each image (Fig. 3).

Figures 3a and 3b show the topographic images and the corresponding current images (C-AFM in air) of two successive scanned zones for the LSMO sample (1st zone inside the dashed lines). We clearly observe a decrease of the current (a factor 10) after the second scan at 0.1V (fig. 1b). We repeated a third time and again observed an additional decrease by a factor 10 (see Fig. S5 in the supporting information). The topography of the sample is not modified (Fig. 3a). These results are in agreement with previous reports.[17-19]

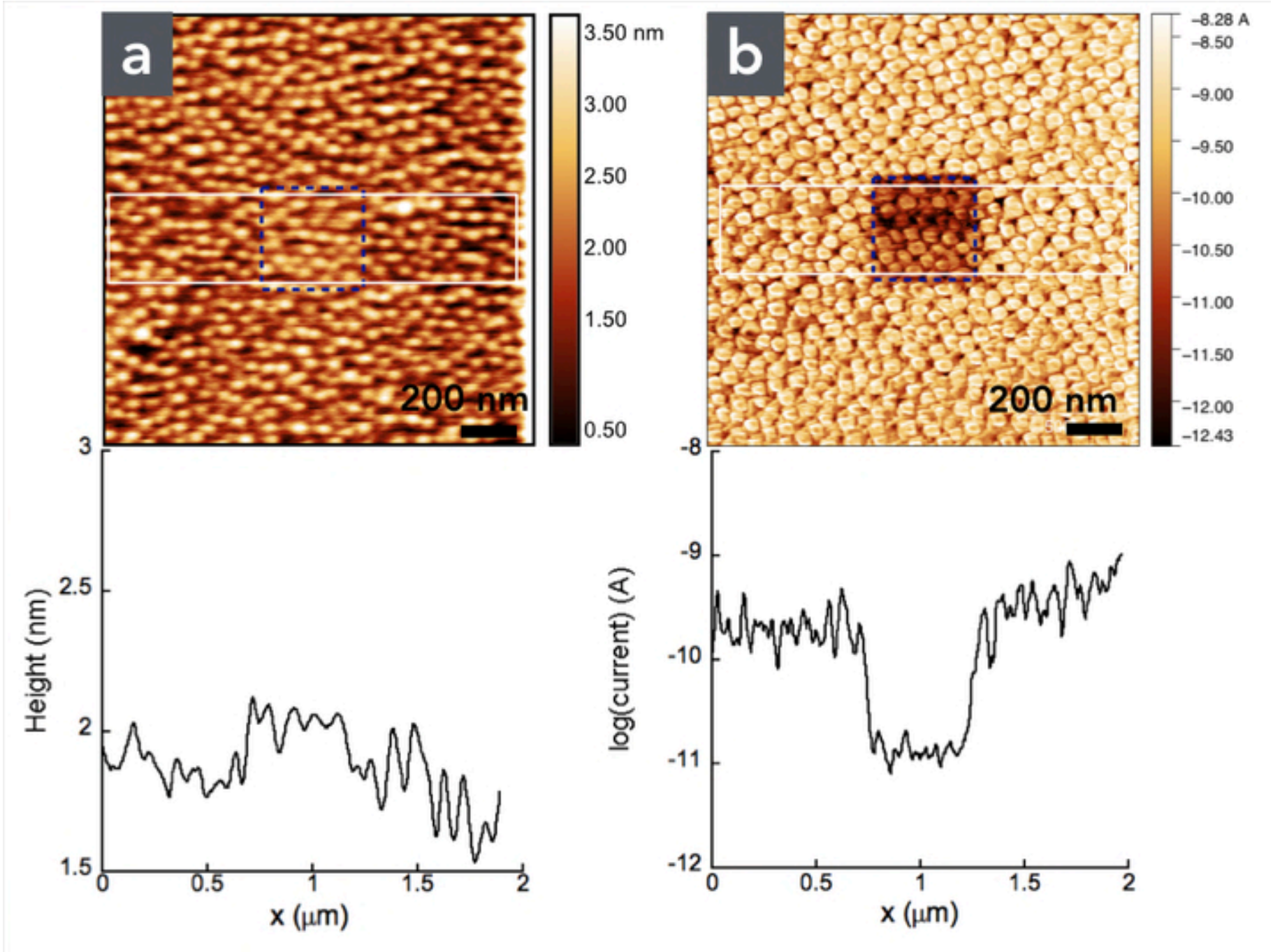

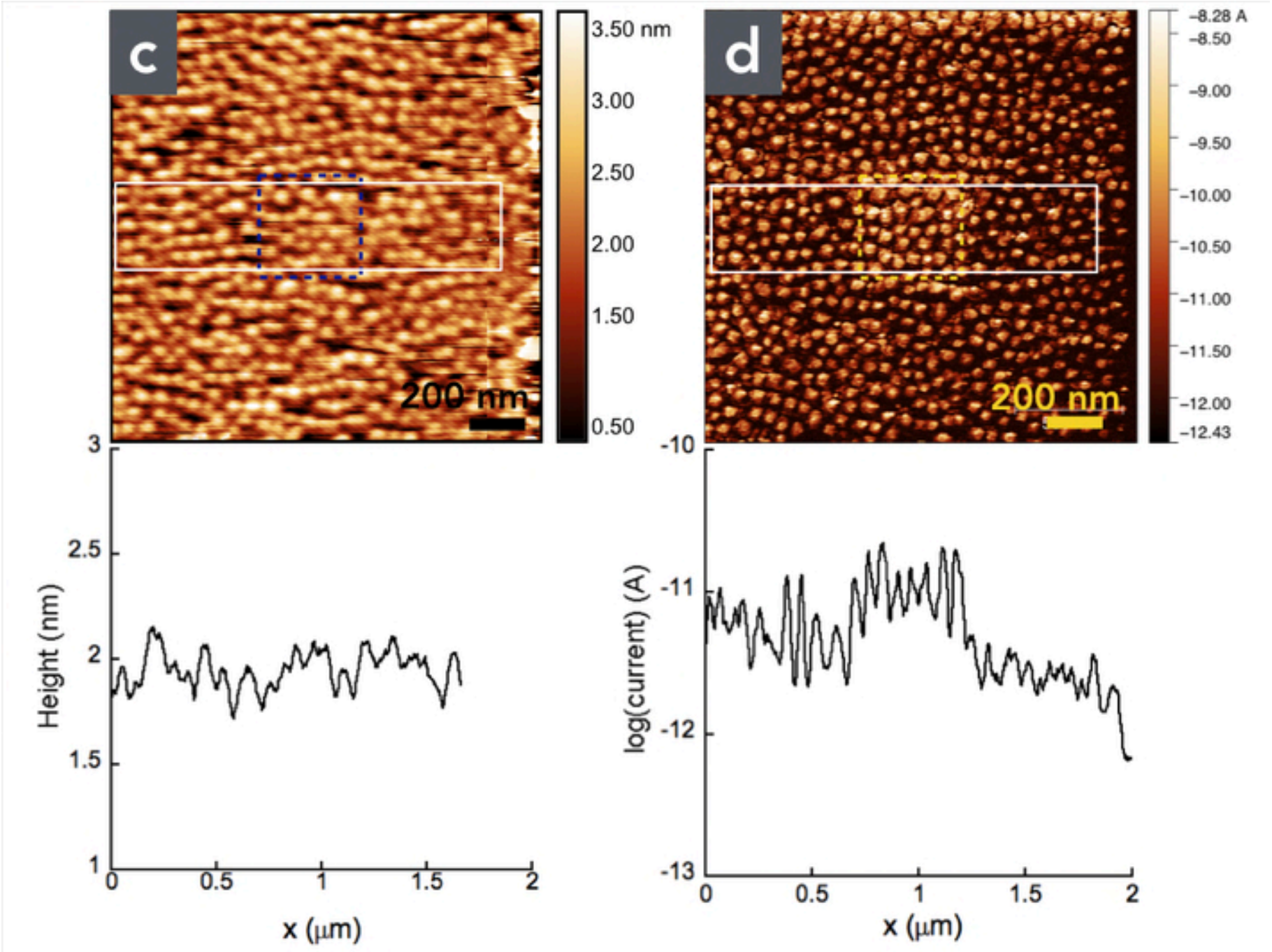

***Figure 3. (a)*** *Topographic AFM (in air) image and* ***(b)*** *current (C-AFM in air) image (log scale) at 0.1 V and loading force 30 nN of the LSMO substrates for two successively scanned zones with increasing dimensions (1st scan 0.5 x 0.5 µm$^2$ marked by the dashed lines, and 2nd scan 2 x 2 µm$^2$). The profiles are averaged on several image lines inside the white rectangles.* ***(c)*** *Topographic AFM (in air) image and* ***(d)*** *current (C-AFM in air) image (log scale) at 0.3 V and loading force 30 nN of the LSMO/DDA samples for two magnifications (1st scan 0.5 x 0.5 µm$^2$ marked by the dashed lines, and 2nd scan 2 x 2 µm$^2$). The profiles are averaged on several image lines inside the white rectangles.*

We also repeated the same experiments on the same sample after the grafting of a DDA SAM to study how the DDA SAM can modify this LSMO switching effect. In that case, the conductance switching is suppressed (Figs. 3c and 3d).
Since oxygen and water can be involved in the LSMO conductance switching behavior,[17-19] the same experiments were conducted in UHV (Fig. 4). Basically, we observed the same effect. On LSMO, the switching ratio is weaker (about 4, fig. 4b, compared to 10). The presence of the SAM still suppresses the LSMO conductance switching (only a very weak ratio of about 1.4 can be distinguished, fig. 4d). We note that a small swelling of the film (about 0.9 nm of the LSMO film thickness, Fig. 4a) is observed, which is similar to previously reported works (but this effect might be an artificial swelling due to changes in the electrostatic interactions between the C-AFM tip and the LSMO surface).[18, 28] This swelling is also strongly reduced in the presence of the SAM (about 0.5 nm, Fig. 4d).

The conductance switching of LSMO was also observed without bias, i.e., only under the mechanical strain induced by the C-AFM tip.[28] The same effect is observed with our samples (see Fig. S6 supporting information).

We discuss several hypotheses to explain the suppression of the LSMO conductance switching in presence of the SAM. We first remind that the origin of this electromechanical conductance switching is ascribed mainly to oxygen vacancy migrations.[17-19, 29, 30] Under the application of a bias voltage, the resulting electric field at the tip/surface interface and in the film induces a modification of the oxygen vacancy concentration, which controls the electron conduction in the film under the C-AFM tip (higher the oxygen vacancy concentration, higher the conductance). In addition, the local compression by the C-AFM tip induces a local flexoelectric effect, modifying the oxygen and electron concentrations in the film.[30] This flexoelectric effect is also combined with a compositional Vegard strain originating from the difference of the electron and ion distribution due to the lattice dilatation (when the concentration of oxygen vacancy[31] is increased, the unit cell volume is also increased, modifying the local electronic properties of the material) as modeled in Ref. 29. It was observed that this conductance switching persists for a long time (hours) due to the slow relaxation process of oxygen vacancies.[30]

Related to these mechanisms, we consider three cases. i) The presence of the SAM between the LSMO surface and the C-AFM tip reduces both the applied electric field - and the current passing - through the LSMO film. However, this reason may be discarded as the dominant mechanism since the conductivity switching was also observed at 0V. ii) The presence of the SAM reduces the effective mechanical strain on the LSMO film, a part of the C-AFM applied pressure being shielded by deformation of the soft SAM. iii) The grafting of the SAM can stabilize the LSMO surface chemistry, thus reducing the change in stoichiometry previously reported[17-19] for these types of mechanical and bias constraints on LSMO films.

We note that other SAMs (e.g. made with alkyl chains or any other molecules) or other capping layers should, in principle, produce the same reduction (or suppression) of this electromechanical conducting switching, but scarce data are available in the literature, to the best of our knowledge. For example, in LSMO/dodecylphosphonic acid SAM/Co magnetic tunnel junctions resistive switching has not been observed for applied bias voltages as large as 2 V.[5] In another report, with a protecting layer of iron oxide nanoparticles (but with a partial surface coverage of around 50%), the switching effect can be still observed, but requiring to apply a very large loading force (> 0.5 µN).[30]

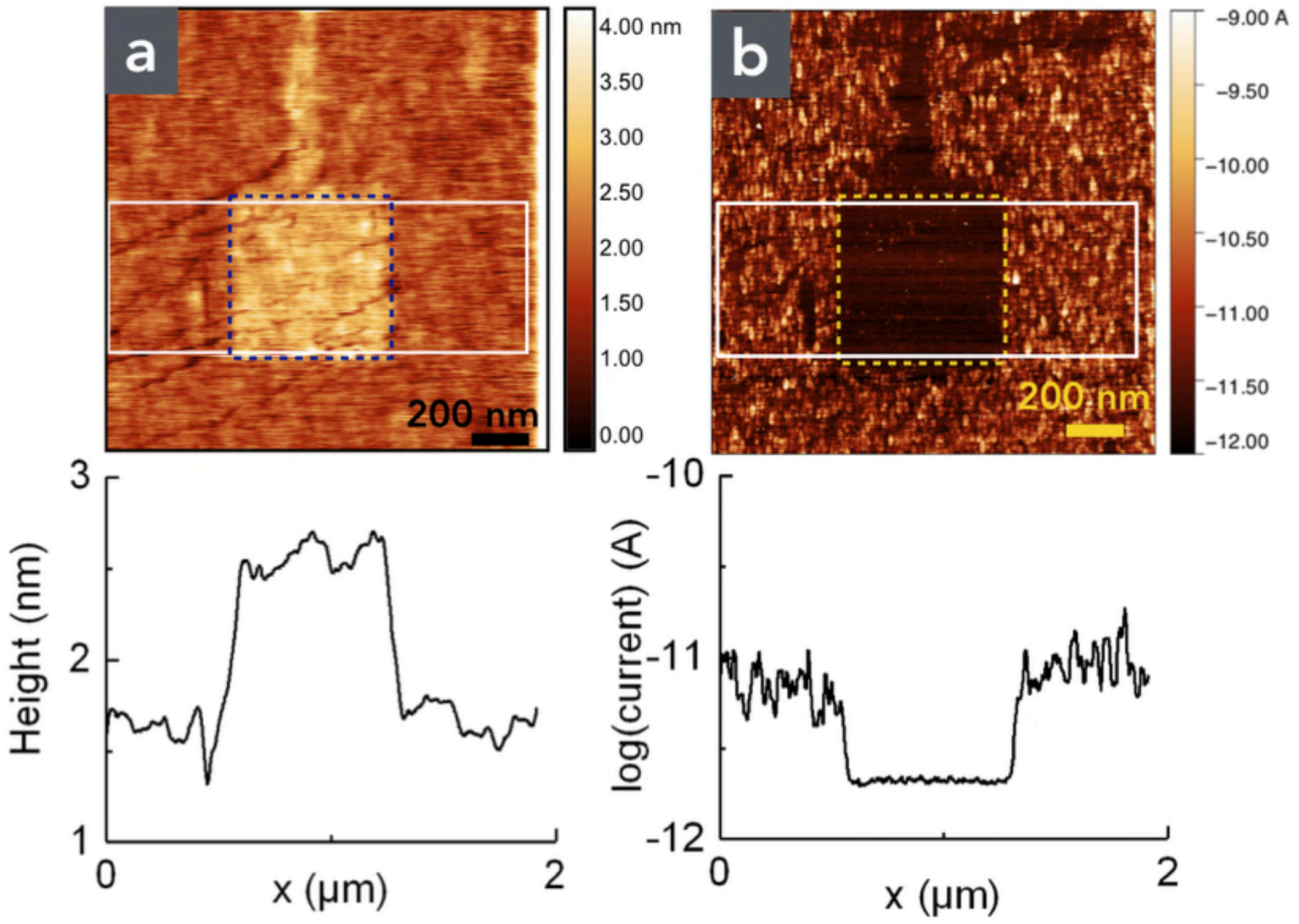

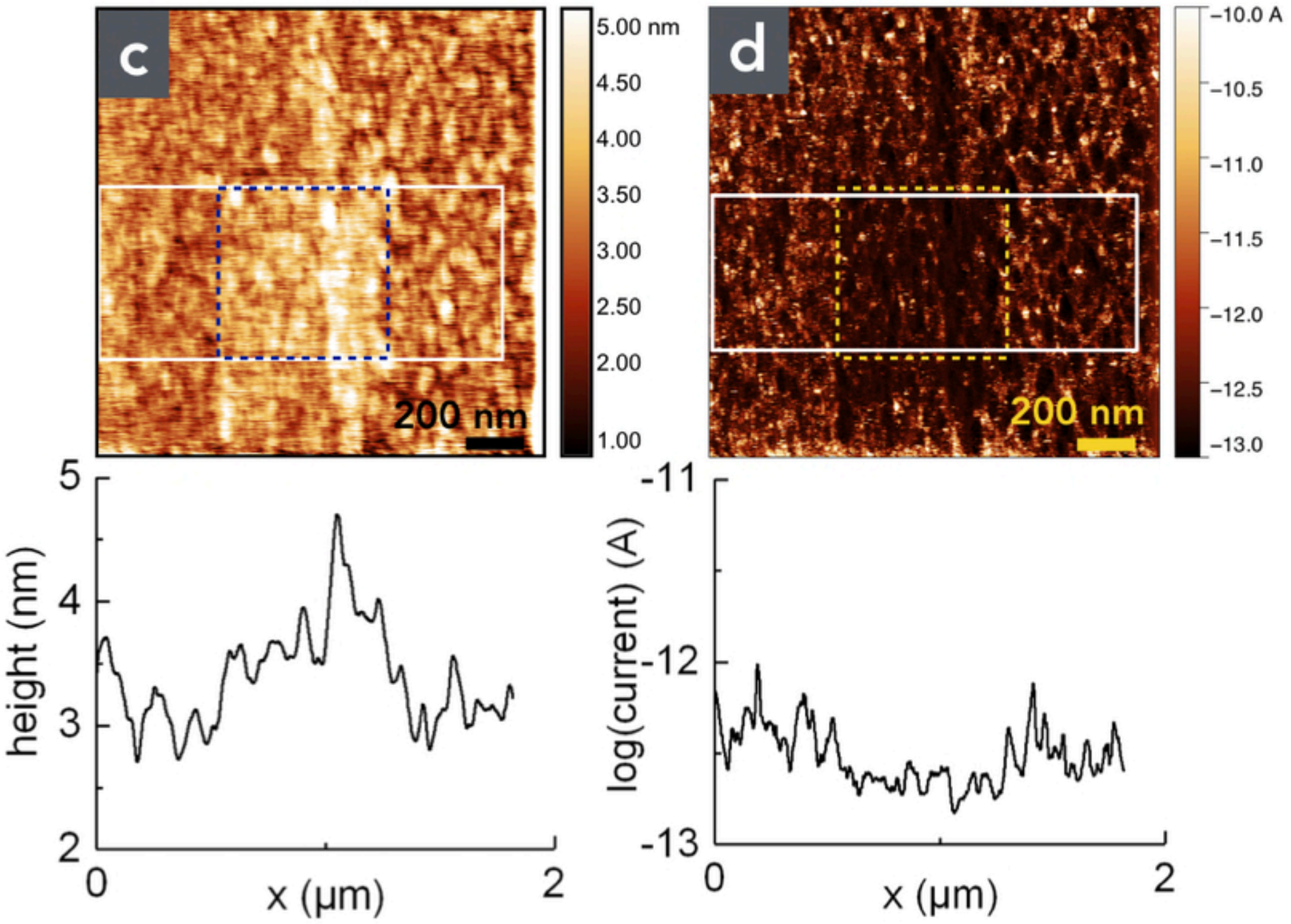

***Figure 4.*** ***(a)*** *Topographic AFM (in UHV) images and* ***(b)*** *current (C-AFM in UHV) images (log scale) at -0.5 V and loading force 30 nN of the LSMO substrates for two successively scanned zones with increasing dimensions (1st scan 0.8 x 0.8 µm$^2$ marked by the dashed lines, and 2nd scan 2 x 2 µm$^2$). The profiles are averaged on several image lines inside the white rectangles.* ***(c)*** *Topographic AFM (in UHV) images and* ***(d)*** *current (C-AFM in UHV) images (log scale) at -1 V and loading force 30 nN of the LSMO/DDA samples for two magnifications (1st scan 0.8 x 0.8 µm$^2$ marked by the dashed lines, and 2nd scan 2 x 2 µm$^2$). The profiles are averaged on several image lines inside the white rectangles.*

The conductance switching of the LSMO being suppressed with the DDA-SAM, we now study the optically induced switching of the DDA. LSMO/DDA samples were irradiated at 365 nm during 30 min to induce the switching to the more conducting closed state DDA-c. Figure 5 shows the 2D current histograms before and after the 365 nm irradiation measured by C-AFM (ambient conditions). Before irradiation, we measured a current distribution with two peaks, P1 with log µ = -11.06 (8.7x10$^{-12}$ A) and P2 with log µ = -10.14 (7.2x10$^{-11}$

A), at V=0.3V. We can ascribe P1 to DDA-o and P2 to DDA-c with a c/o conductance ratio of about 8, in agreement with previous results showing a higher molecular conductance when the diarylethene is in the closed from.[11, 13, 32, 33] From the areas of the two peaks (ratio P1/P2 = 3.7), we deduce that about 80 % of the DDA are in the open form after grafting.

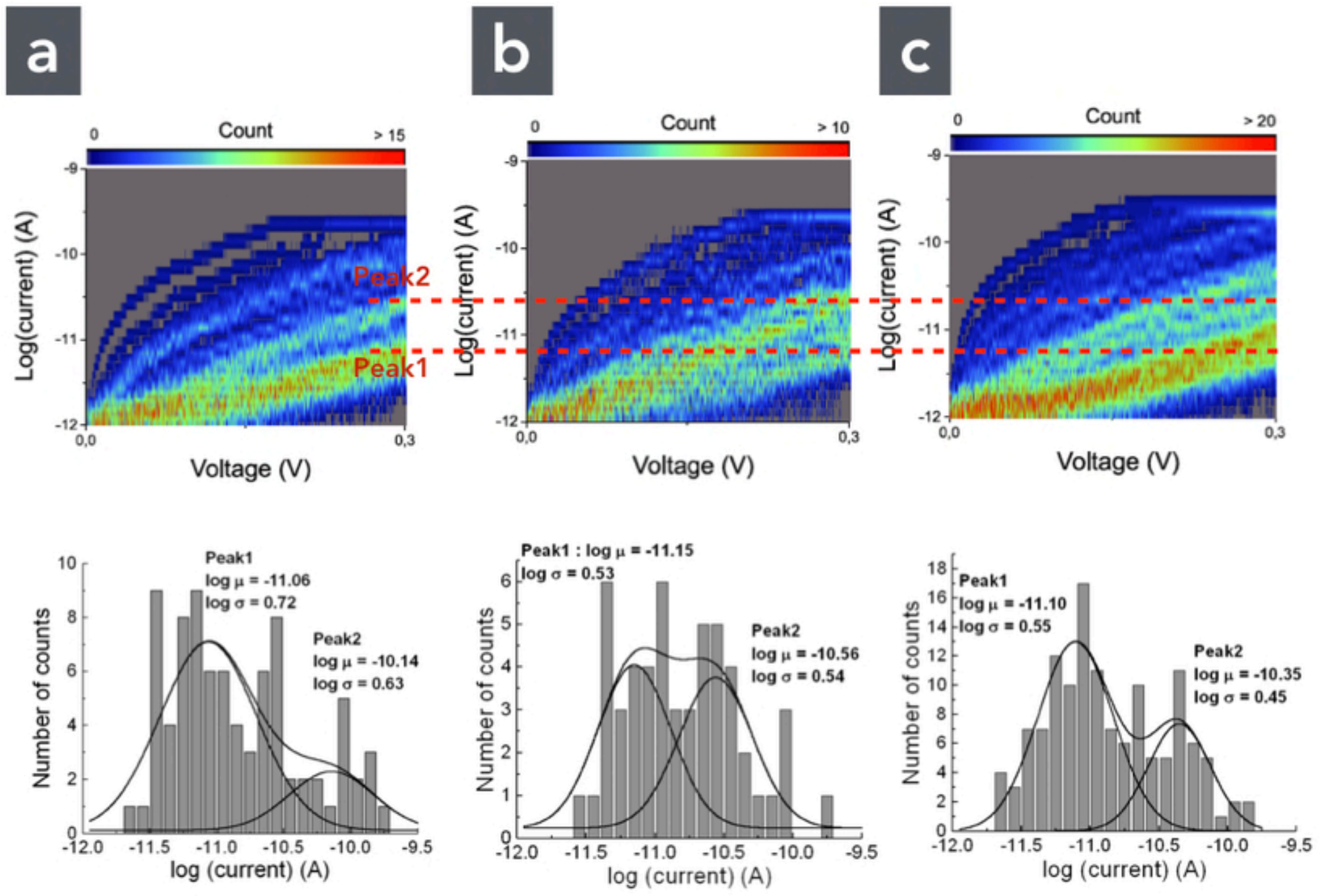

***Figure 5.*** ***(a)*** *2D histogram of current-voltage behavior of the LSMO/DDA after grafting the DDA on the LSMO substrate (85 I-V traces) and 1D histogram at 0.3 V.* ***(b)*** *2D histogram of current-voltage behavior of the LSMO/DDA after 30 min irradiation at 365 nm (56 I-V traces) and 1D histogram at 0.3 V.* ***(c)*** *2D histogram of current-voltage behavior of the LSMO/DDA after 12 h in the dark (148 I-V traces) and 1D histogram at 0.3 V. All measurements by C-AFM (in air), loading force 30 nN. Note that the saturation of the current amplifier (plateau at log(current)=-9.5 in the 2D histograms) has been removed on the 1D histograms. The black lines are the fits with a log-normal distribution. The fit parameters, log-mean current (log μ) and log-standard deviation (log σ) are given in the figures. Note that the relatively small number of I-V traces in these histograms (while around thousands were acquired, see details in the supporting information) is*

*due to the fact that a majority of the traces showed currents <$2x10^{-12}$ A, close to the sensitivity and noise level of our apparatus, and were removed for clarity. Nevertheless, a number of counts >20-30 is reasonably sufficient to a significant statistical analysis of C-AFM measurements on SAMs.*[34]

After the UV irradiation (Fig. 5b), a weak increase of the average current is seen on the 2D histograms. The doted red lines visualize the current level of the maximum of counts at 0.3 V for the two peaks P1 and P2 attributed to the DDA-o and DDA-c, respectively. From the 1D histogram, we still have two peaks (P1 : log µ = -11.15 ($7.1x10^{-12}$ A) , P2 : log µ = -10.56 ($2.7x10^{-11}$ A)), but we observed a relative decrease of the amplitude of P1 and an increase of the amplitude of P2 (P1/P2 = 1.07) indicating a partial switching of the DDA-o to DDA-c. Then, after 12h in the dark, Fig. 5c shows that a majority (68 %, P1/P2=2.2) of the molecules have switched back to the open form (P1 : log µ = -11.10 ($7.9x10^{-12}$ A) , P2 : log µ = -10.35 ($4.5x10^{-11}$ A)). We conclude that a partial switching between the open and closed forms has occurred. Under almost the same irradiation conditions (power, duration), the DDA switching behavior was observed for 3 different samples (out of a total of 6) - see more data for the two other "switching" samples in the supporting information (Fig. S7 and S8). No switching was observed for two samples. In an attempt to foster the switching feature, a peculiar "one shot" switching was observed for another LSMO/DDA sample irradiated for longer periods of time (6h) (Fig. S9 in the supporting information). In that case, we observed a non-reversible decrease of the current after this long UV irradiation. A possible explanation would be that long times (6h) irradiation at 365 nm can induce the blocked form of the diarylethene.[35] After a subsequent irradiation of 6h at 470 nm, no change was observed. Moreover, in that case, the switching behavior of the DDA molecules can also be hindered by a persistent photoresistivity (PPR) - Fig S10 in the supporting information, i.e., increase of LSMO resistance (a factor about 15) after a long time UV light irradiation. This effect has also been recently observed in macroscopic (4 probes) measurements of LSMO films on STO, and this effect was ascribed to photoinduced spin disorder in the film.[36] Thus, it is likely that this peculiar "one-shot" conductance switching of the LSMO/DDA (Fig. S9) under a long time UV irradiation is due to the combination of both the blocked form of the DDA and the photoresistivity effect of the LSMO substrate.

## Conclusion.

In conclusion, we have extended the phosphonic acid route for the grafting on LSMO of functional molecules such as molecular optical switch (diarylethene derivatives). Compact SAMs of diarylethene derivatives (dithienylethene diphosphonic acid) are formed on LSMO and the conductance of LSMO/DDA reduced by about 3 decades with the DDA mainly in its open form (compared to the conductance of the LSMO substrate). The presence of the DDA SAM suppresses (or strongly reduces) the known conductance switching of the LSMO substrate that is induced by mechanical and/or bias constraints under C-AFM measurements. Finally, only a partial light-induced conductance switching between the open and closed forms of the DDA is observed for the LSMO/DDA samples (close/open conductance ratio of about 8). However, in the case of long-time exposition to UV light, this feature can be partly hindered by a larger, non-reversible, conductance decrease (a factor of about 15) of the LSMO electrode. This results calls for the design and synthesis of diarylethene derivatives with a higher (100 and larger) conductance ratio between the close and open forms to improve the performances of the LSMO/DDA molecular junctions.

## Author contributions.

L.T and D. G. fabricated the SAMs, and D. G. synthesized and characterized the DDA molecules. B.Q and E.J fabricated the LSMO substrates with R.M. and P.S. L.T. carried out the experiments during his PhD thesis under the supervision of T.M. and S.L. D.V. and S.L. supervised the project at IEMN, R.M. and P.S. supervised the project at CNRS/Thales. D.V. wrote the paper with L.T., T.M. and S.L., all the authors discussed the results and contributed to the redaction of the manuscript.

## ORCID

R. Mattana: 0000-0002-8815-6434

D. Vuillaume: 0000-0002-3362-1669

T. Melin: 0000-0003-1777-3512

S. Lenfant: 0000-0002-6857-8752

## Conflict of interest.

The authors declare no competing financial interests.

## Acknowledgements.

This work was financially supported by an ANR grant "spinfun" (n° ANR-17-CE24-0004). This work was supported by the "IDI" project funded by the IDEX Paris-Saclay, ANR-11-IDEX-0003-02. We thank C. van Dyck (U. of Alberta, Canada), V. Diez Cabanes and J. Cornil (LCNM, U. Mons, Belgium) for theoretical discussions on the electronic properties of diarylethene derivatives. We thank S. Godey and D. Deresmes (IEMN-CNRS) for advises and collaborations with the UHV C-AFM experiments, and J.L. Caudron (IEMN-CNRS) for XPS measurements.

# Conductance switching at the nanoscale of diarylethene derivatives self-assembled monolayers on $La_{0.7}Sr_{0.3}MnO_3$

L. Thomas,[1] D. Guérin,[1] B. Quinard,[2] E. Jacquet,[2] R. Mattana,[2] P. Seneor,[2]
D. Vuillaume,[1,*] T. Mélin[1] and S. Lenfant.[1,*]

*1. Institute for Electronics Microelectronics and Nanotechnology (IEMN), CNRS, Univ. Lille, 59652 Villeneuve d'Ascq, France.*

*2. Unité Mixte de Physique CNRS/Thales, CNRS, Thales, University Paris Sud, Université Paris Saclay, 91767 Palaiseau, France.*

* Corresponding author : stephane.lenfant@iemn.fr ; dominique.vuillaume@iemn.fr

## Supporting information

### Synthesis of dithienylethene diphosphonic acid (DDA).

Dibromide **1** then diphosphonate **2** precursors were prepared according to the method of Feringa et al.[1] We followed the method of Reisinger et al.[2] to synthesize the dithienylethene diphosphonic acid (DDA).

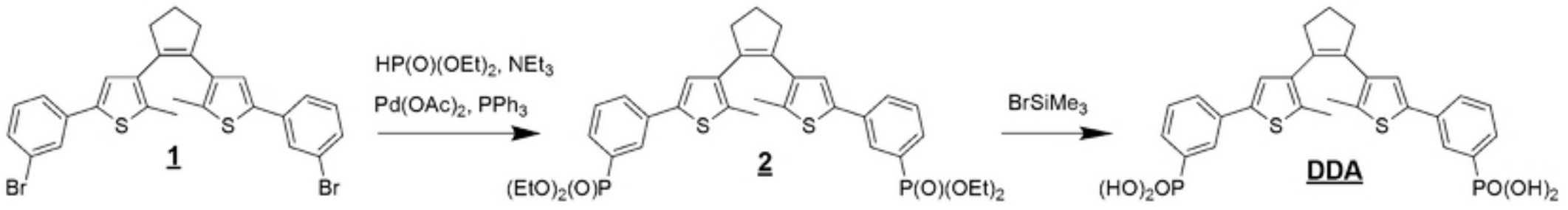

Reaction was performed under an inert atmosphere in oven-dried glassware and protected from UV light. A solution of diphosphonate **2** (250 mg, 0.36 mmol) in dry acetonitrile (20 mL) was treated with trimethylsilyl bromide (0.75 mL, 4.6 mmol). The mixture was stirred overnight at room temperature under nitrogen. Volatile compounds were removed under vacuum then 20 mL of methanol was added then the solution was stirred overnight. After evaporation of the solvent, the residue was kept at 60°C/0.1 mbar for 3 h to remove traces of solvents and silylated compounds. A control by TLC on $SiO_2$ (AcOEt/hexane 95:5) of the obtained yellow oil indicated the total disappearance of diphosphonate **2**. Addition of $CH_2Cl_2$ to the oil provided DDA as a thin white solid that was isolated by centrifugation (177 mg, 84%). DDA was purified by additional cleaning with $CH_2Cl_2$, centrifugation then drying under vacuum. Yield: 84 %. $^1H$ NMR δ (300.13 MHz, $CD_3OD$) : 1.99 (6H, s, thiophene-$CH_3$), 2.10 (2H, quint, J = 7.4 Hz, cyclopentene-$CH_2$), 2.85 (4H, t, J = 7.4 Hz, cyclopentene-$CH_2$), 7.16 (2H, s, thiophene-H), 7.40-7.46 (2H, m, phenyl-H), 7.61-7.68 (4H, m, phenyl-H), 7.94 (2H, d, J = 14.2 Hz, phenyl-H). $^{13}C$ {$^1H$, $^{31}P$} NMR δ (100.67 MHz, d6-DMSO) : 14.06, 22.41, 38.06, 124.79, 126.78, 127.21, 129.02, 129.25, 133.45, 134.18, 134.31, 135.02, 136.85, 138.30. $^{31}P$ {$^1H$} NMR δ (162.05 MHz, d6-DMSO) : 12.94. NSI-HRMS: m/z $[M+H]^+$ calcd. for $C_{27}H_{26}O_6P_2S_2$: 573.07188, found : 573.07134.

## NMR spectra of DDA :

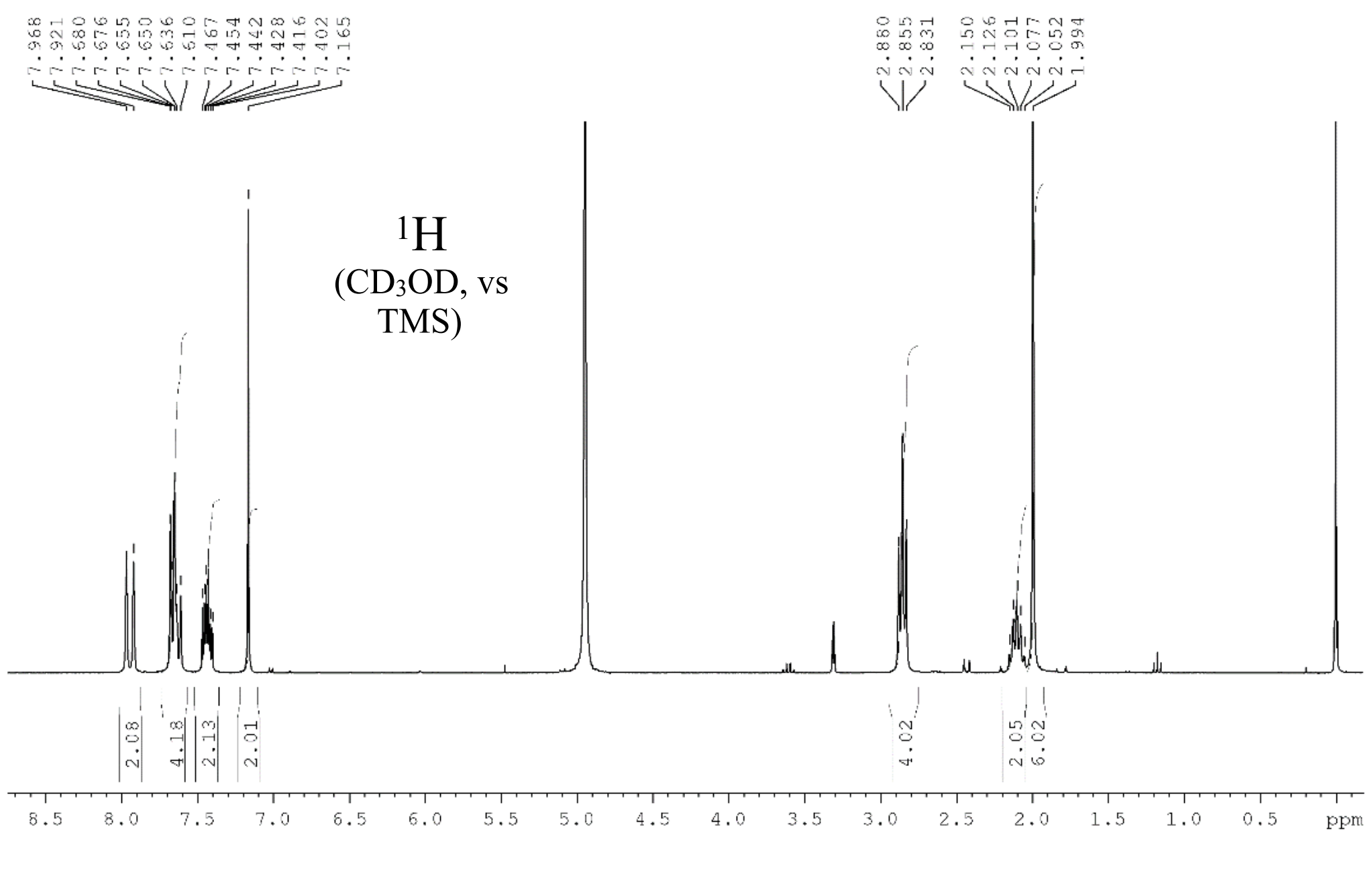

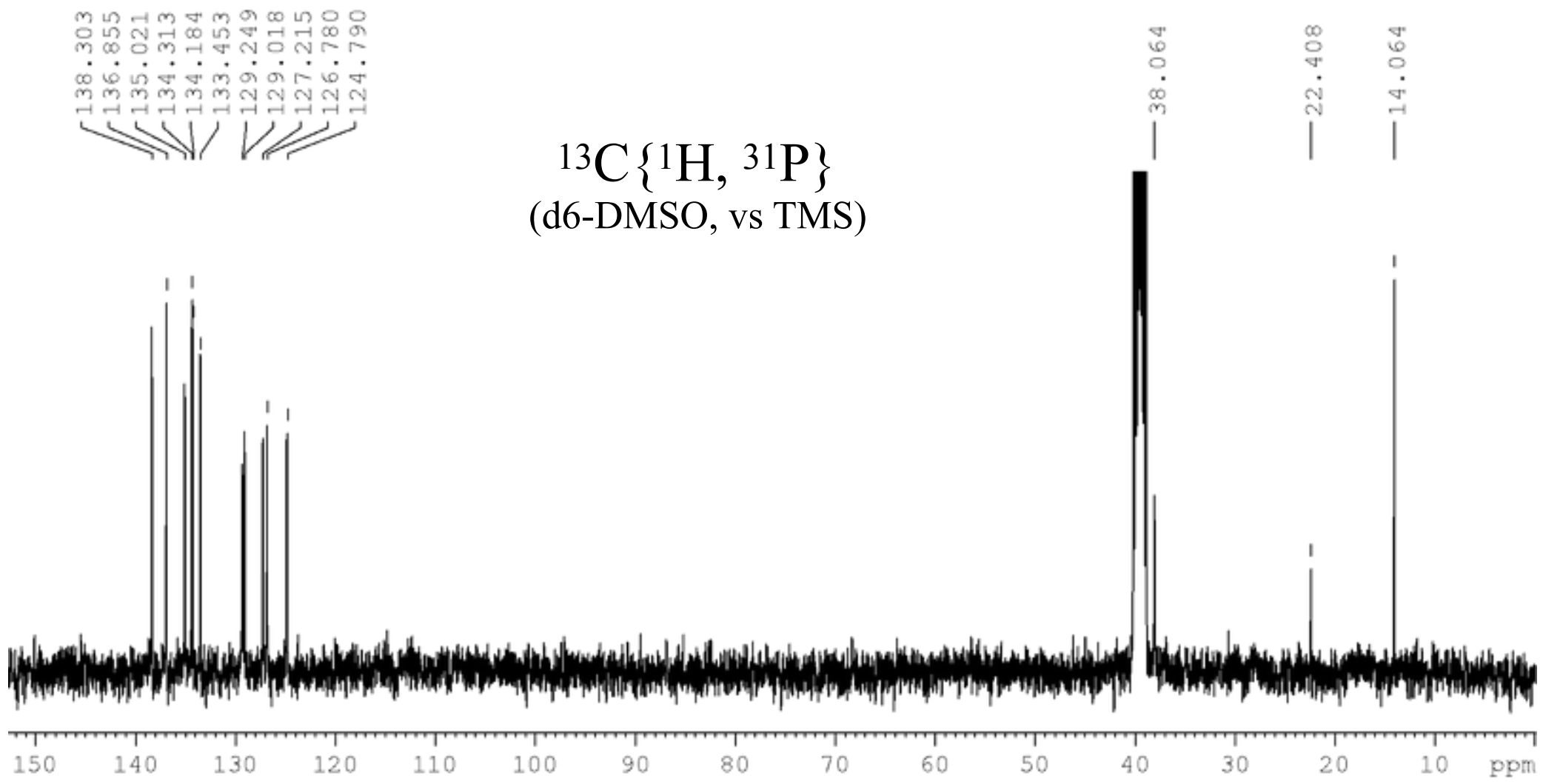
138.303
136.855
135.021
134.313
134.184
133.453
129.249
129.018
127.215
126.780
124.790
38.064
22.408
14.064
13C{1H, 31P}
(d6-DMSO, vs TMS)
150 140 130 120 110 100 90 80 70 60 50 40 30 20 10 ppm

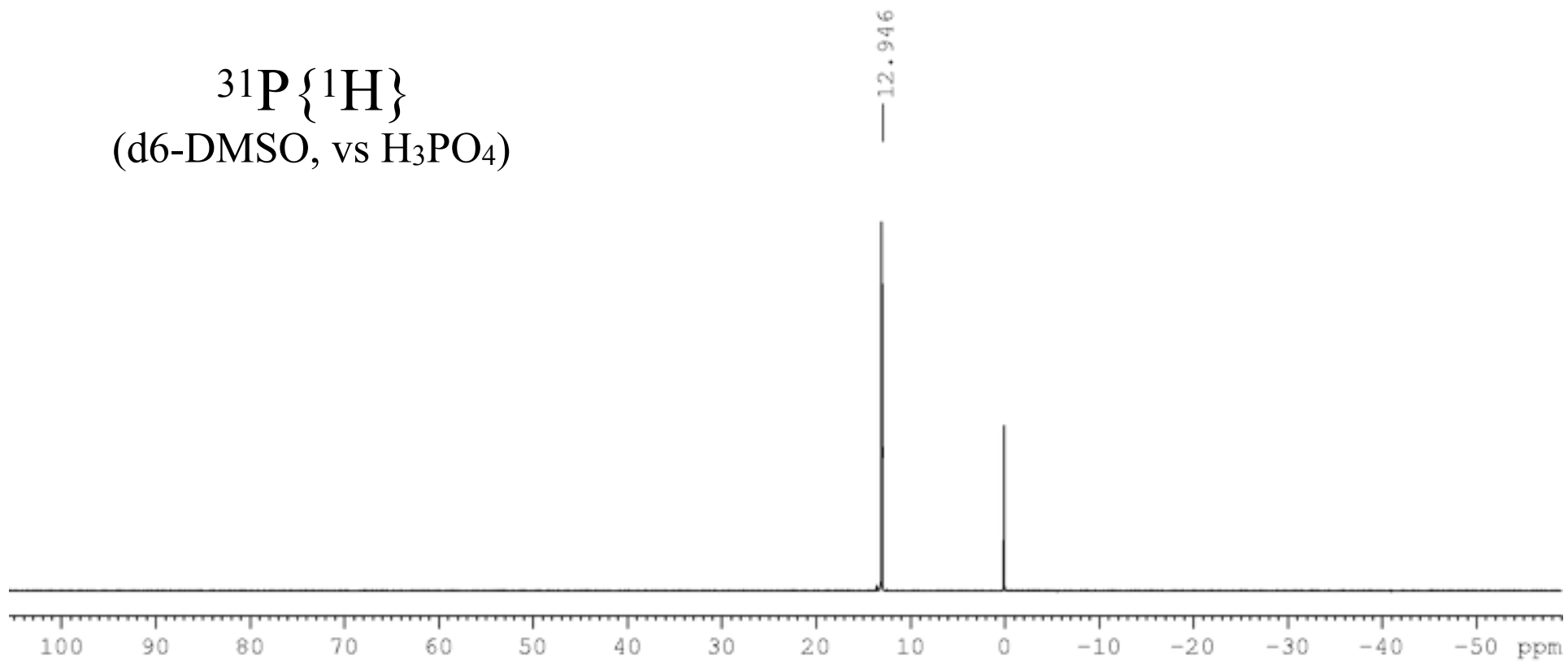
12.946
31P{1H}
(d6-DMSO, vs H3PO4)
100 90 80 70 60 50 40 30 20 10 0 −10 −20 −30 −40 −50 ppm

HRMS spectrum :

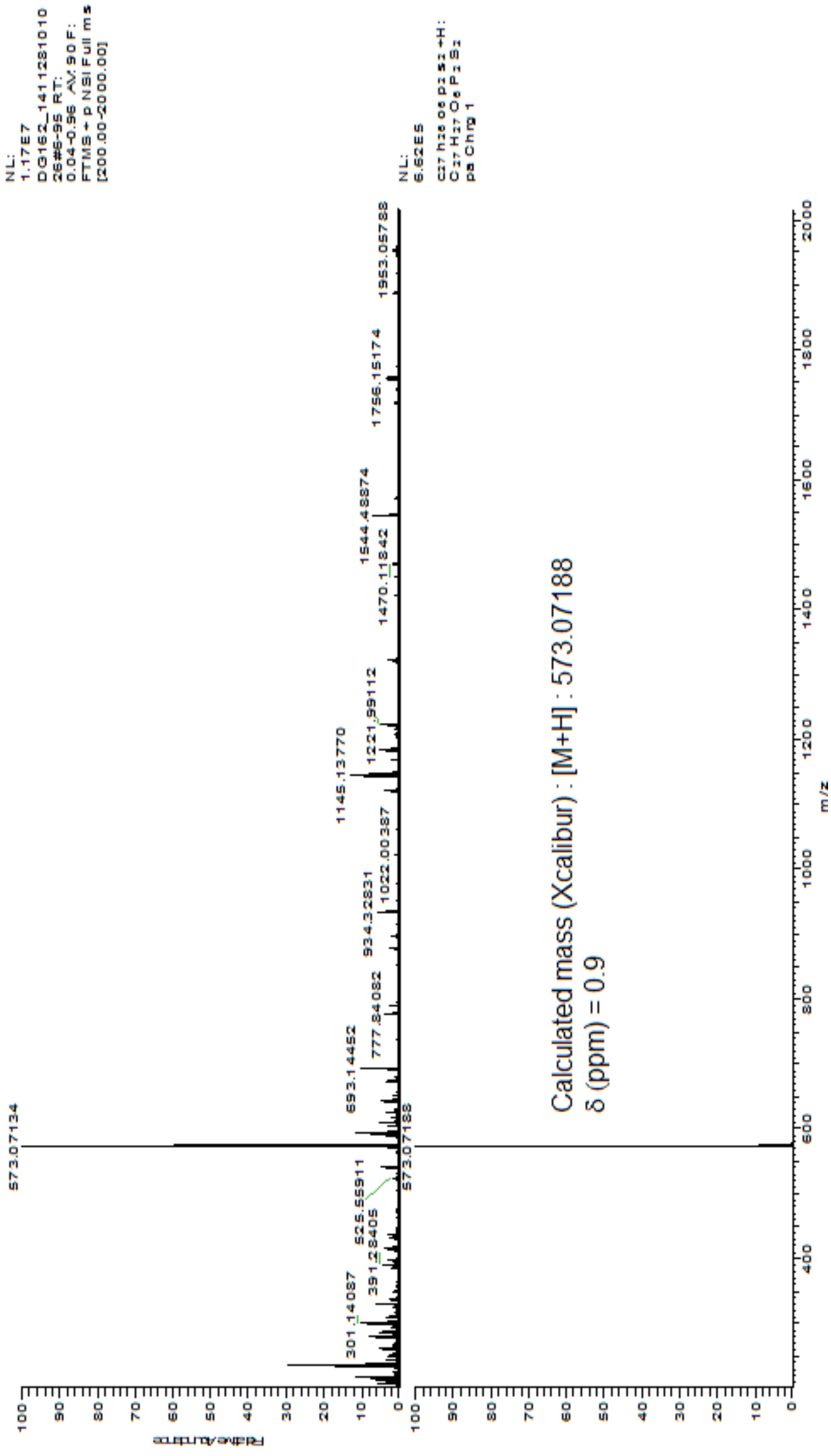

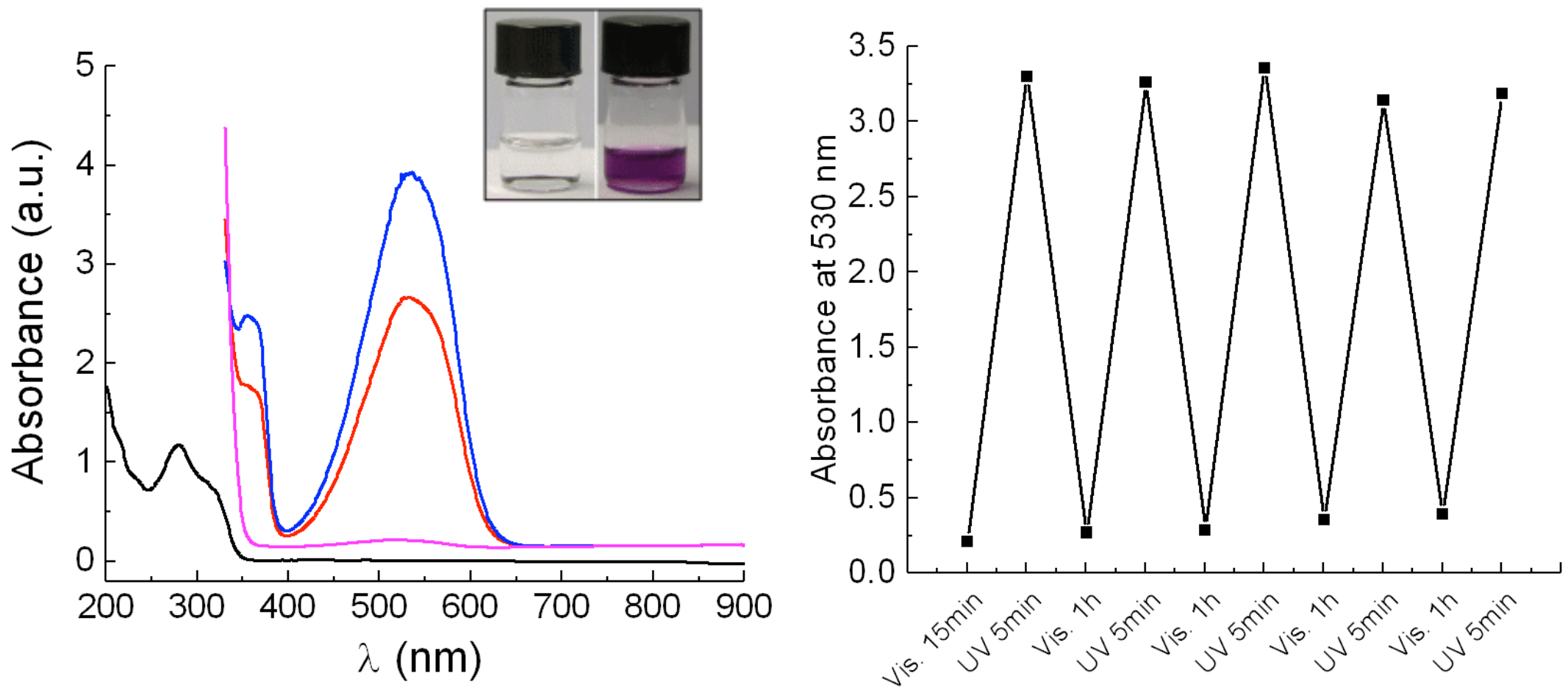

***Figure S1***. ***(a)*** *UV-vis absorption of DDA in ethanol: reference before irradiation (black line), after irradiation at 365 nm for 2 min (red line), for 5 min (blue line) and after irradiation 15 min in visible (pink line). The peak at 530 nm is the fingerprint of the DDA-c form. Inset : solution before (left) and after the 365 nm irradiation (right).* ***(b)*** *Reversible switching of the absorbance at 530 nm (main peak) under several cycles of UV and visible light irradiations. Note that we tried a similar UV-vis experiment on the DDA SAM on LSMO. Albeit we can observe a very small peak at around 530 nm, the signal was too weak (about $10^{-3}$ with respect to the LSMO contribution) to resolve any signature of the DDA switching.*

## Ellipsometry.

We recorded spectroscopic ellipsometry data in the visible range using an UVISEL (Jobin Yvon Horiba) spectroscopic ellipsometer equipped with DeltaPsi 2 data analysis software. The system acquired a spectrum ranging from 2 to 4.5 eV (corresponding to 300 to 750 nm) with intervals of 0.1 eV (or 15 nm). Data were taken at an angle of incidence of 70°, and the compensator was set at 45°. Data were fitted by a regression analysis to a film-on-substrate model as described by their thickness and their complex refractive indexes. First, a background for the LSMO substrate before monolayer deposition was recorded. Secondly, after the monolayer deposition, we used a 2-layer model (substrate/SAM) to fit the measured data and to determine the SAM thickness. We employed the previously measured optical properties of the LSMO substrate (background), and we fixed the refractive index of the organic monolayer at 1.50. The usual values in the literature for the refractive index of organic monolayers are in the range 1.45−1.50.[3] We can notice that a change from 1.50 to 1.55 would result in less than 1 Å error for a thickness less than 30 Å. We estimated the accuracy of the SAM thickness measurements at ± 2 Å.

## Contact angle measurements.

We measured the water contact angle with a remote-computer controlled goniometer system (DIGIDROP by GBX, France). We deposited a drop (10-30 μL) of deionized water (18MΩ.cm$^{-1}$) on the surface and the projected image was acquired and stored by the computer. Contact angles were extracted by a contrast contour image analysis software. These angles were determined few seconds after the application of the drop. These measurements were carried out in a clean room (ISO 6) where the relative humidity (50%) and the temperature (22°C) are controlled. The precision with these measurements are ± 2°.

## XPS.

XPS was performed with a Physical Electronics 5600 spectrometer fitted in an UHV chamber with a residual pressure of $2\times10^{-10}$ Torr. High resolution spectra were recorded with a monochromatic Al K$\alpha$ X-ray source ($h\nu$ = 1486.6 eV), a detection angle of 45° as referenced to the sample surface, an analyzer entrance slit width of 400 μm and with an analyzer pass energy of 12 eV. Semi-quantitative analysis was completed after standard background subtraction according to Shirley's method.[4] Peaks were decomposed by using Voigt functions.

## AFM and C-AFM.

Atomic force microscopy (topography) and conducting atomic force microscopy (C-AFM) were performed in air (of under a flux of dry $N_2$) (ICON, Bruker), using a tip probe in platinum/iridium or platinum. The tip loading force on the surface was to 30 nN. Albeit larger than the usual loading force (2-5 nN) used for C-AFM on SAMs, this value is below the limit of about 60-70 nN at which the SAMs start to suffer from severe degradations. For example, a detailed study (Ref. 5) showed a limited strain-induced deformation of the monolayer (≲ 0.3 nm). The same conclusion was  confirmed by our own study comparing mechanical and electrical properties of alkylthiol SAMs on flat Au surfaces and tiny Au nanodots.[6] In addition, we note that imaging several times the same area of the SAMs (e.g. as in Figs. 3c, 4c) showed no significant degradation of the SAMs. We have also chosen this value of the loading force, since the currents on the DDA/LSMO SAMs are low and would not be measured (or only a much weaker number, thus degrading the statistical analysis) using a weaker loading force. The topographic and current images are recorded simultaneously. To measure the current-voltage (I-V) curves and the current histograms, a square grid of 10×10 or 20x20 was defined with a pitch of 50 or 100 nm. At each point, the I-V curve is acquired leading to the measurements of 100 or 400 I-V traces per grid. This process was repeated several times at different places on the sample, and up to thousands of I-V traces were used to construct the current-voltage histograms. The bias was applied on the LSMO substrate and the tip was grounded through the input of the current amplifier. Note that around 0V the currents are very weak and in many cases at the limit of detection (0.1 pA).

C-AFM in UHV ($10^{-9}$ - $10^{-11}$ mbar) was performed using a VT-SPM (Variable temperature scanning probe microscope, Scienta Omicron) using a PtIr coated tip (SCM-PIC-V2). The tip loading force was set to 30 nN. Current-voltage spectroscopy was performed on 20×20 grids with a pitch of 100 nm. Grids were spaced a few mm away from each other. Under UHV, the bias was applied on the probe and the sample was grounded.

## UV-vis irradiations.

For the light exposures, an optical fiber, with a 400 µm diameter, was brought near the samples. We used 2 power LED from Thorlabs, with the following characteristics: i) 365 nm, bandwidth of 10 nm, for the open-to-closed isomerization, ii) 470 nm, bandwidth of 20 nm, for the close-to-open isomerization. The light power values were measured with a calibrated Optical Power and Energy meter PM200 (Thorlabs). They were 10.2 mW, 8.6 mW, on the devices at the output of the optical fiber for the 365 and 470 nm sources, respectively. These powers gave the following power densities of 4.4 $mW/cm^2$ (UV) and 3 $mW/cm^2$ (blue) with the C-AFM in air, and 7.6 $mW/cm^2$ (UV) and 6 $mW/cm^2$ (blue) with the C-AFM in UHV, depending on the exact geometries of the set-up. For the switching in solution (Fig. S1), we used a chromatography UV lamp (Vilbert Lourmat, power density 2.7 $mW/cm^2$), and a large band halogen lamp (Schott KL 2500 LCD, power density 74 $mW/cm^2$) for the irradiation in visible.

## XPS results.

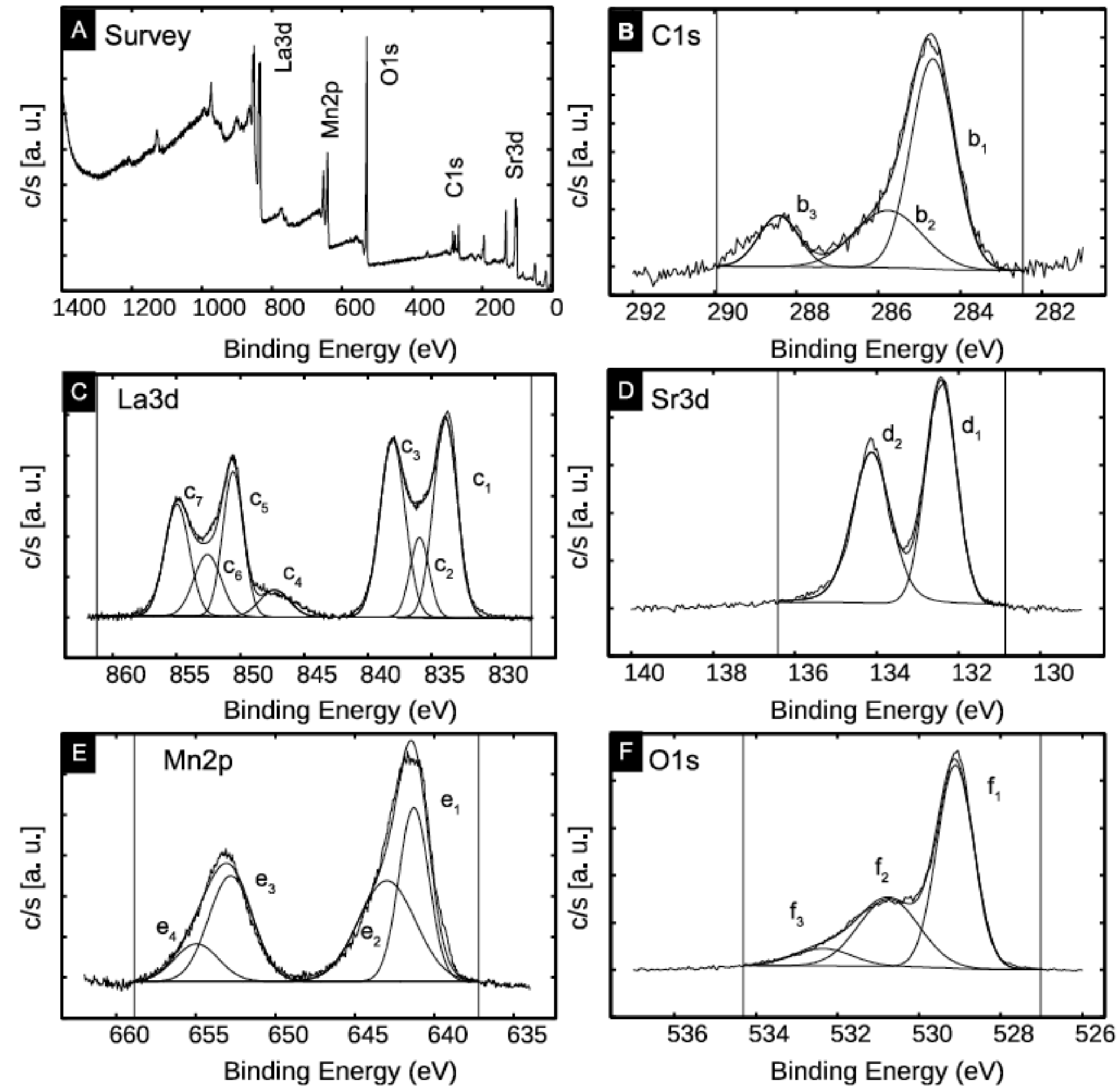

***Figure S2.*** *XPS spectra of LSMO substrate.* ***(a)*** *survey spectrum, and high resolution spectra of* ***(b)*** *C1s,* ***(c)*** *La3d,* ***(d)*** *Sr3d,* ***(e)*** *Mn2p and* ***(f)*** *O1s contributions. The component of the deconvolution, peak area and chemical assignments are analyzed in Table S1.*

| Spectrum | Peak | Identification | Energy [eV] | Area [a. u.] | ASF [a. u.] | Area / ASF [a. u.] | 0,7 × Ratio La |
|---|---|---|---|---|---|---|---|
| C1s | b1 | C-C | 284.7 | 994 | -- | -- | |
| C1s | b2 | C-O-C | 285.8 | 405 | -- | -- | |
| C1s | b3 | O-C=O | 288.4 | 215 | -- | -- | |
| La3d 5/2 | c1 | La (LSMO) | 833.9 | 11466 | -- | -- | |
| La3d 5/2 | c2 | satellite | 835.9 | 3697 | -- | -- | |
| La3d 5/2 | c3 | satellite | 838.1 | 10854 | -- | -- | |
| La3d 3/2 | c4 | Auger, plasmon | 847.3 | 1952 | -- | -- | |
| La3d 3/2 | c5 | La (LSMO) | 850.6 | 7291 | -- | -- | |
| La3d 3/2 | c6 | satellite | 852.6 | 4099 | -- | -- | |
| La3d 3/2 | c7 | satellite | 855 | 6708 | -- | -- | |
| La3d 5/2 + 3/2 | (Sum ci) - c4 | -- | -- | 44115 | 7.708 | 5723 | 0.7 |
| Sr3d 5/2 | d1 | Sr (LSMO) | 132.4 | 2213 | -- | -- | -- |
| Sr3d 3/2 | d2 | Sr (LSMO) | 134.1 | 2036 | -- | -- | -- |
| Sr3d 5/2 + 3/2 | Sum di | -- | -- | 4249 | 1.578 | 2693 | 0.33 |
| Mn2p 3/2 | e1 | Mn3+ | 641.3 | 5080 | -- | -- | -- |
| Mn2p 3/2 | e2 | Mn4+ | 643 | 5556 | -- | -- | -- |
| Mn2p 1/2 | e3 | Mn3+ | 652.8 | 4505 | -- | -- | -- |
| Mn2p 1/2 | e4 | Mn4+ | 655 | 1599 | -- | -- | -- |
| Mn2p 3/2 + 1/2 | Sum ei | -- | -- | 16740 | 2.42 | 6917 | 0.85 |
| O1s | f1 | Mn oxide | 529.1 | 9740 | -- | -- | -- |
| O1s | f2 | La oxide | 530.7 | 5326 | -- | -- | -- |
| O1s | f3 | Sr oxide | 532.4 | 1331 | -- | -- | -- |
| O1s | Sum fi | -- | -- | 16397 | 0.711 | 23062 | 2.82 |

***Table S1.*** *Deconvolution of the XPS peaks and chemical assignment, peak energy, peak area, atomic sensitivity factor (ASF), ASF corrected contribution of each chemical element, and deduced stoichiometry of LSMO assuming a weight 0.7 for La. Peak assignation from Refs. 7, 8.*

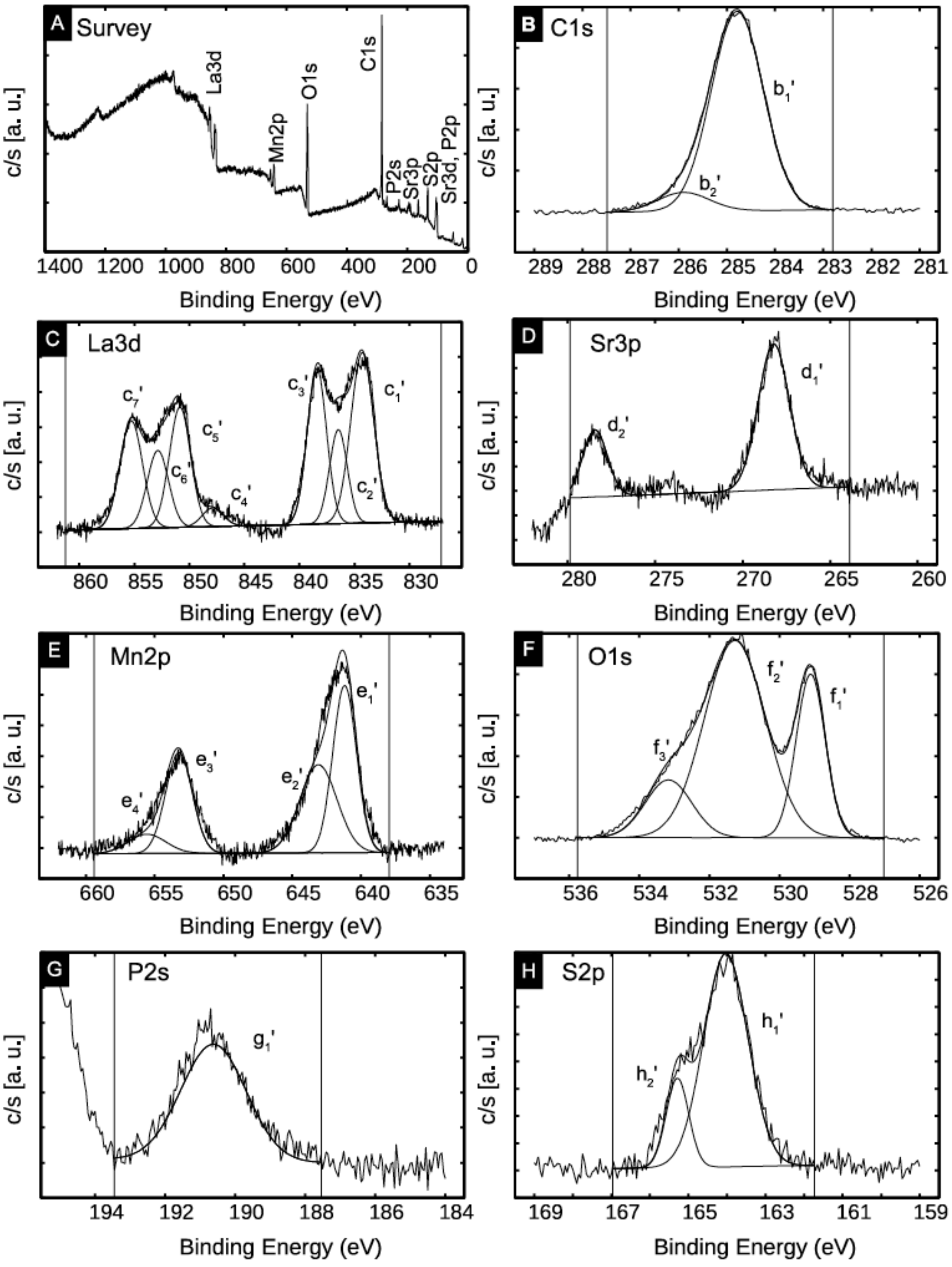

***Figure S3.*** *XPS spectra of LSMO/DDA sample.* ***(a)*** *survey spectrum, and high resolution spectra of* ***(b)*** *C1s,* ***(c)*** *La3d,* ***(d)*** *Sr3d,* ***(e)*** *Mn2p,* ***(f)*** *O1s,* ***(g)*** *P2s and* ***(h)*** *S2p contributions. The component of the deconvolution, peak area and chemical assignments are analyzed in Table S2.*

| Spectrum | Peak | Identification | Energy [eV] | Area [a. u.] | ASF [a. u.] | Area / ASF | 0,7 × Ratio La | 2,0 × Ratio S |
|---|---|---|---|---|---|---|---|---|
| C1s | b1' | C-C | 284.8 | 6638 | -- | -- | | |
| C1s | b1'-b1 | C-C | -- | 5644 | 0.296 | 19067 | -- | 31.78 |
| C1s | b2' | C-O-C | 285.9 | 593 | -- | -- | -- | -- |
| La3d 5/2 | c1' | La (LSMO) | 834.3 | 2543 | -- | -- | -- | -- |
| La3d 5/2 | c2' | satellite | 836.5 | 1084 | -- | -- | -- | -- |
| La3d 5/2 | c3' | satellite | 838.4 | 1936 | -- | -- | -- | -- |
| La3d 3/2 | c4' | Auger, plasmon | 847.9 | 255 | -- | -- | -- | -- |
| La3d 3/2 | c5' | La (LSMO) | 850.8 | 1662 | -- | -- | -- | -- |
| La3d 3/2 | c6' | satellite | 852.9 | 1070 | -- | -- | -- | -- |
| La3d 3/2 | c7' | satellite | 855.3 | 1583 | -- | -- | -- | -- |
| La3d 5/2 + 3/2 | (Sum ci') - c4' | -- | -- | 9878 | 7.708 | 1282 | 0.7 | -- |
| Sr3p 3/2 | d1' | Sr oxide | 286.2 | 668 | -- | -- | -- | -- |
| Sr3p 1/2 | d2' | Sr oxide | 278.4 | 241 | -- | -- | -- | -- |
| Sr3p 5/2 + 3/2 | Sum di' | -- | -- | 909 | 1.536 | 592 | 0.32 | -- |
| Mn2p 3/2 | e1' | Mn3+ | 641.2 | 1177 | -- | -- | -- | -- |
| Mn2p 3/2 | e2' | Mn4+ | 643.1 | 972 | -- | -- | -- | -- |
| Mn2p 1/2 | e3' | Mn3+ | 653.2 | 796 | -- | -- | -- | -- |
| Mn2p 1/2 | e4' | Mn4+ | 655.6 | 217 | -- | -- | -- | -- |
| Mn2p 3/2 + 1/2 | Sum ei' | -- | -- | 3162 | 2.42 | 1307 | 0.71 | -- |
| O1s | f1' | Mn oxide | 529.1 | 2081 | -- | -- | -- | -- |
| O1s | f2' | La oxide, PO(OH)2 | 531.3 | 5286 | -- | -- | -- | -- |
| O1s | f3' | Sr oxide, PO(OH)2 | 533.2 | 1171 | -- | -- | -- | -- |
| O1s | Sum fi' | -- | -- | 8538 | 0.711 | 12008 | -- | -- |
| P2s | g1' | PO(OH)2 | 190.6 | 388 | 0.355 | 1093 | -- | 1.82 |
| S2p 3/2 | h1' | C-S-C | 164 | 564 | -- | -- | | -- |
| S2p 1/2 | h2' | C-S-C | 165.3 | 120 | -- | -- | | -- |
| S2p 3/2 + 1/2 | Sum hi' | -- | | 684 | 0.57 | 1200 | | 2 |

***Table S2***. *Deconvolution of the XPS peaks and chemical assignment, peak energy, peak area, atomic sensitivity factor (ASF), ASF corrected contribution of each chemical element, deduced LSMO stoichiometry assuming a weight 0.7 for La, and DDA stoichiometry. Peak assignation from Refs. 7, 8.*

## Variability of the LSMO substrates.

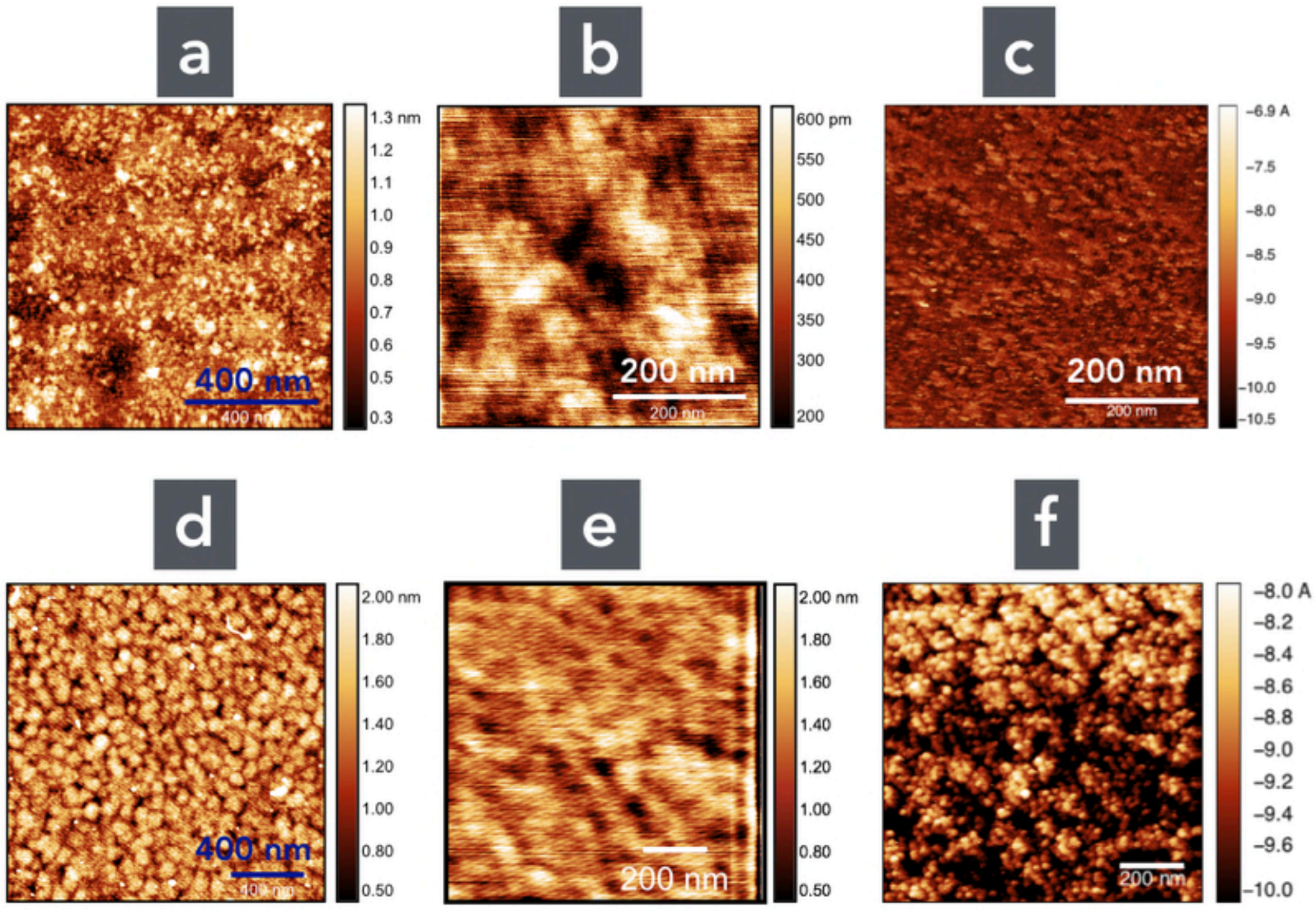

***Figure S4.*** *Sample #1:* ***(a,b)*** *Topographic AFM images (in air) (1x1 $\mu m^2$ and 0.5x0.5 $\mu m^2$, respectively).* ***(c)*** *Current (C-AFM) images (log scale) at 2V, 0.5x0.5 $\mu m^2$. Sample #2:* ***(d,e)*** *Topographic AFM images (in air) (1.6x1.6 $\mu m^2$ and 1x1 $\mu m^2$, respectively).* ***(f)*** *Current (C-AFM) images (log scale) at 0.5 V , 1x1 $\mu m^2$.*

**Repeatability of the stress effect on the LSMO sample.**

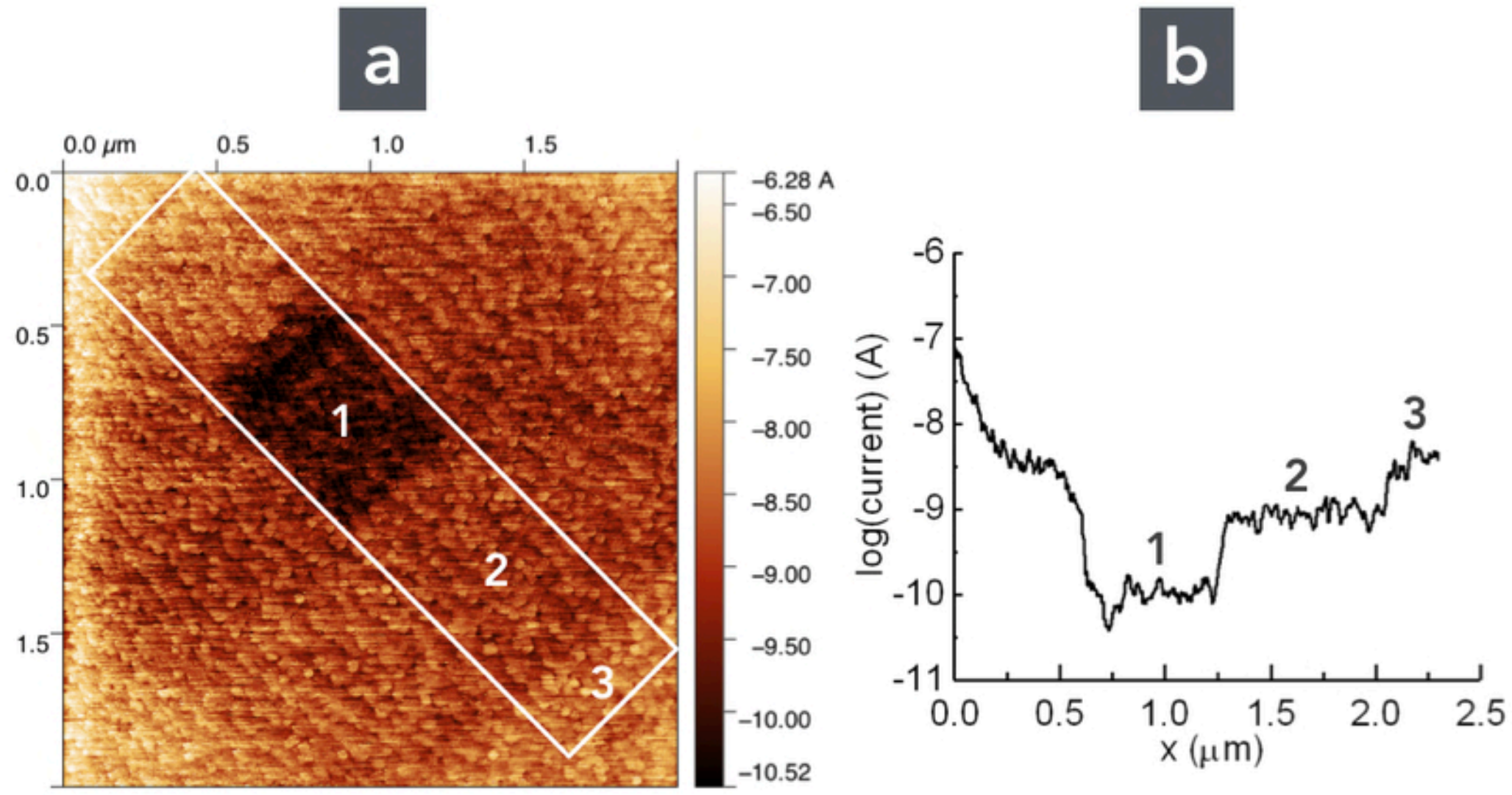

***Figure S5.*** *(a) C-AFM image (log scale, at 0.5V and loading force 30 nN) for 3 successive zones with increasing areas (1: 0.5x0.5 μm$^2$, 2: 2x2 μm$^2$ and 3: 2x2 μm$^2$ rotated by 45°. (b) Current profile averaged on several image lines inside the white rectangle.*

## Conductance switching of the LSMO at 0 V bias.

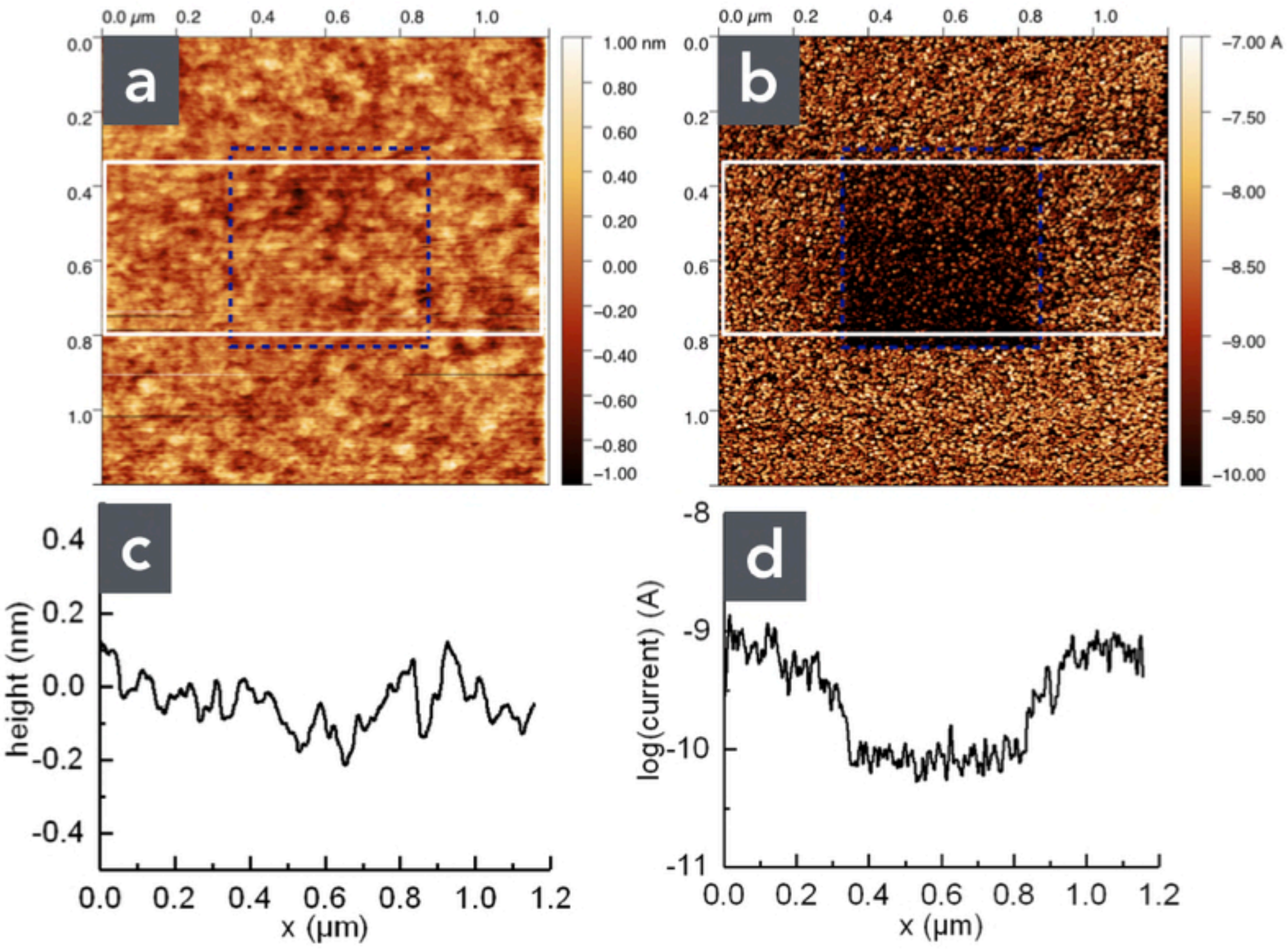

***Figure S6.*** ***(a)*** *Topographic AFM images and* (***b)*** *C-AFM images (current in log scale) of two zones scanned successively by increasing the scanned area dimensions (0.5x0.5 µm² at 0 V and loading force of 30 nN marked by the dashed line, and 1.2x1.2 µm² at 0.1 V loading force 30 nN).* ***(c,d)*** *height and current profile averaged on several image lines inside the white rectangles.*

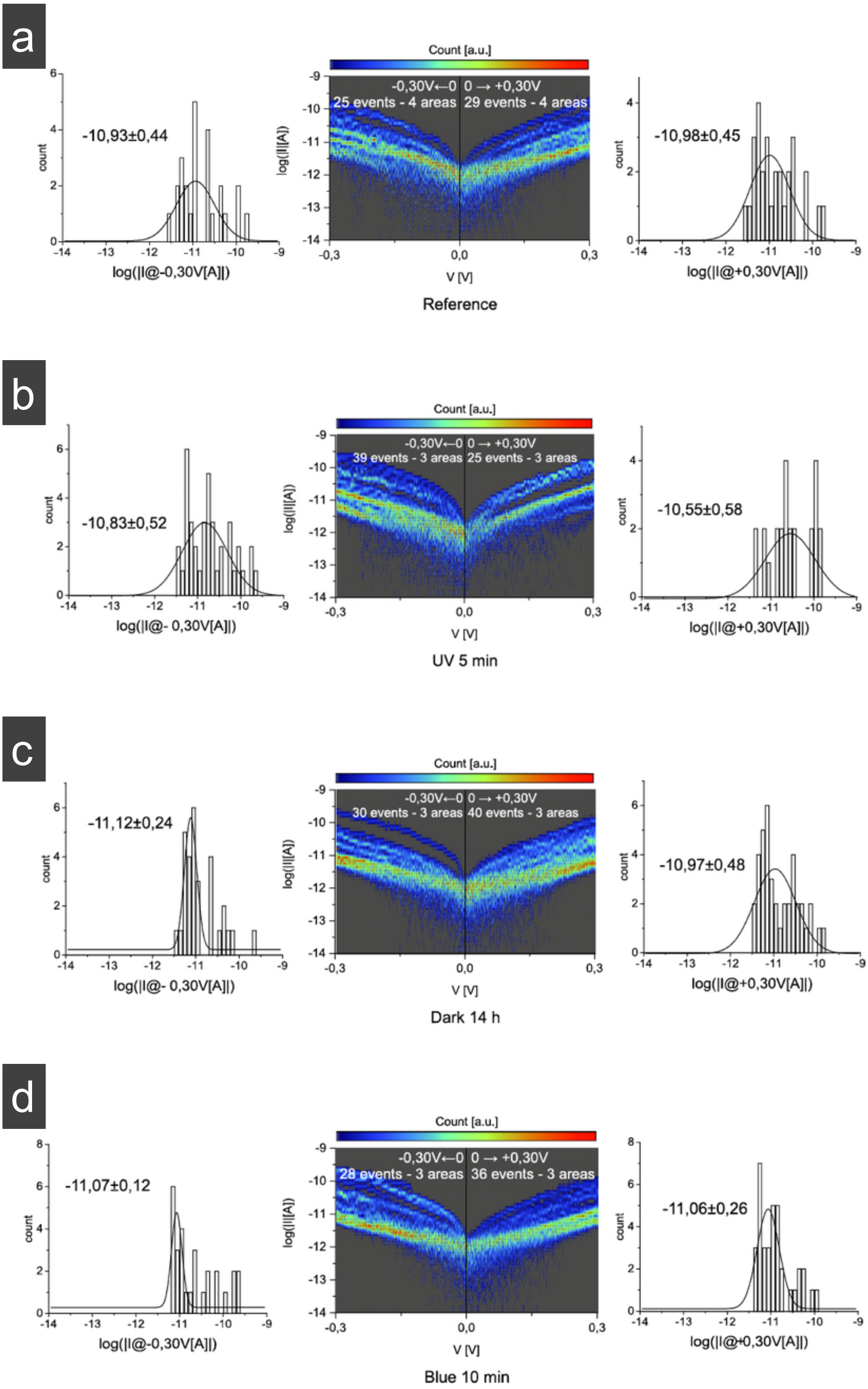
a
-10,93±0,44
count
log(|I@-0,30V[A]|)
Count [a.u.]
-0,30V←0 0 → +0,30V
25 events - 4 areas 29 events - 4 areas
log(|I|[A])
V [V]
Reference
-10,98±0,45
log(|I@+0,30V[A]|)
b
-10,83±0,52
log(|I@- 0,30V[A]|)
-0,30V←0 0 → +0,30V
39 events - 3 areas 25 events - 3 areas
UV 5 min
-10,55±0,58
log(|I@+0,30V[A]|)
c
-11,12±0,24
log(|I@- 0,30V[A]|)
-0,30V←0 0 → +0,30V
30 events - 3 areas 40 events - 3 areas
Dark 14 h
-10,97±0,48
log(|I@+0,30V[A]|)
d
-11,07±0,12
log(|I@-0,30V[A]|)
-0,30V←0 0 → +0,30V
28 events - 3 areas 36 events - 3 areas
Blue 10 min
-11,06±0,26
log(|I@+0,30V[A]|)

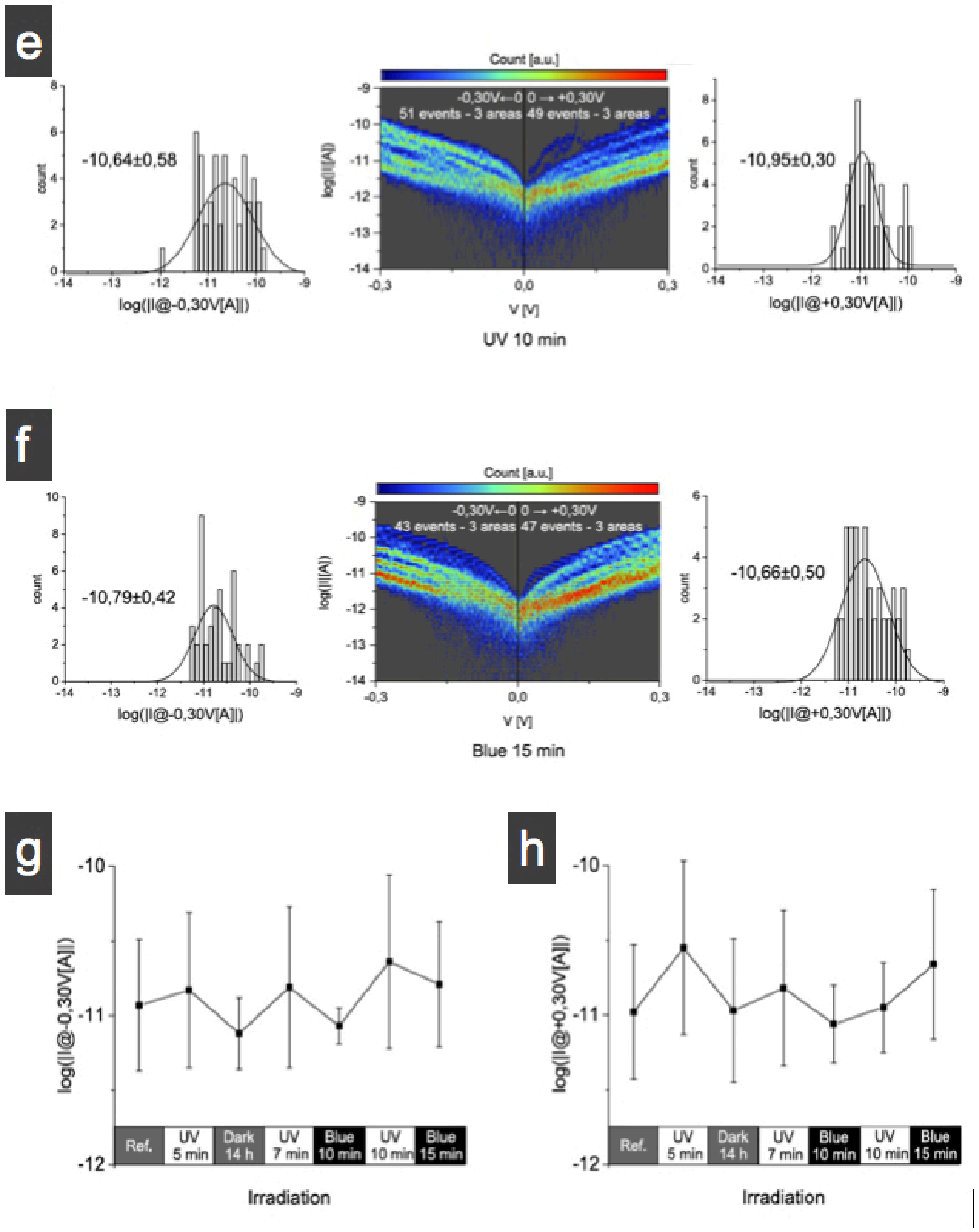

***Figure S7.*** ***(a)*** *2D histogram of current-voltage measurements of the LSMO/DDA samples (middle figure) and 1D histograms at -0.3 V and +0.3 V (left and right figures). The number of measurements on different zones are indicated on the figure, as well as the values of log μ.* ***(b)*** *Same data after 5 min irradiation at 365 nm.* ***(c)*** *After 14h in the dark (slow relaxation to the open form).* ***(d)*** *After 10 min blue irradiation at 470 nm.* ***(e)*** *After 10 min irradiation at 365 nm. (****f)*** *After 15 min blue irradiation at 470 nm. (****g,h)*** *Evolution of the log-mean current at -0.3 and*

*0.3 V, respectively, under the successive stimulations. All C-AFM measurements under a flux of $N_2$ and loading force of 30 nN.*

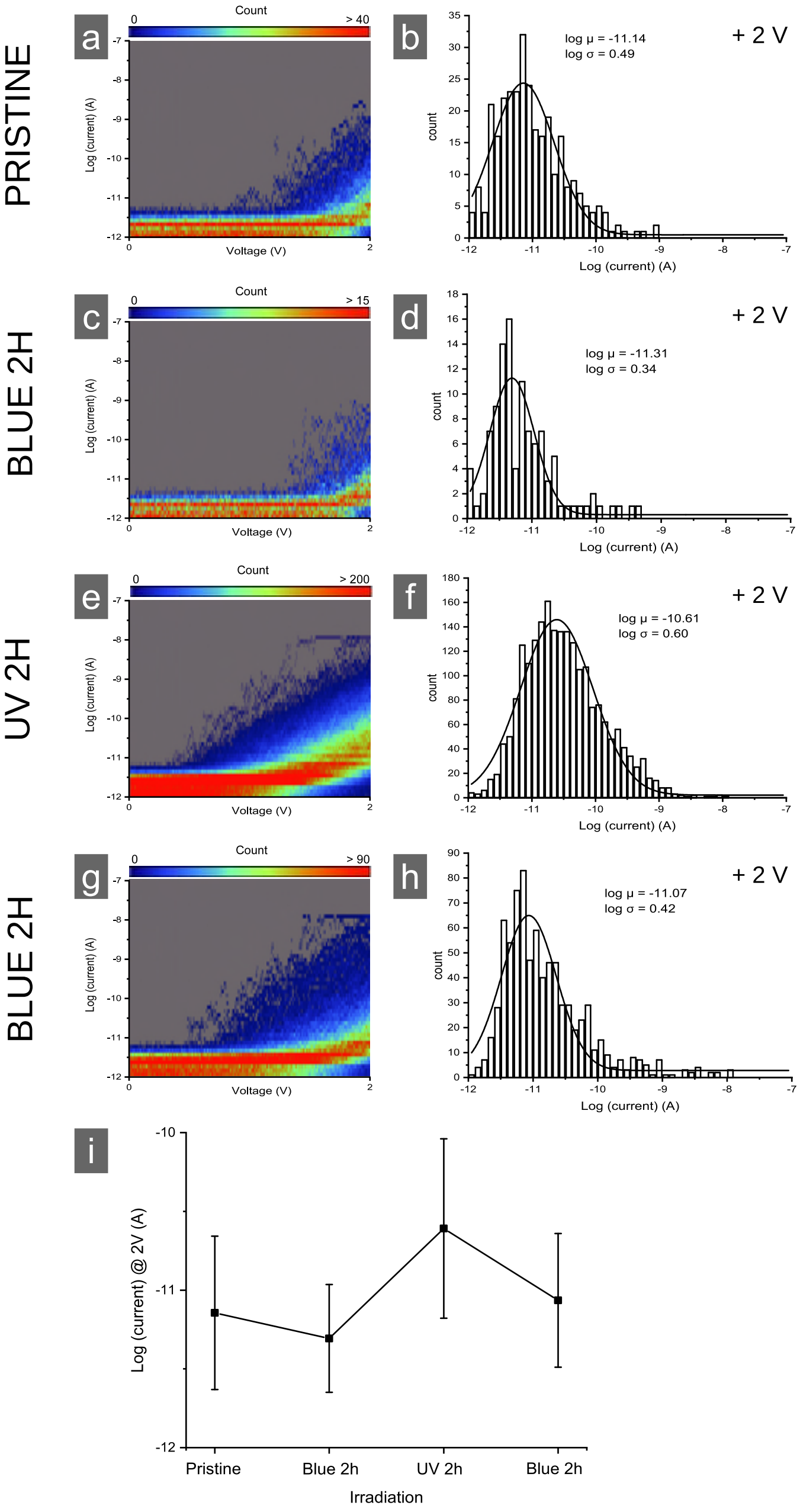

***Figure S8. (a)** 2D histogram of current-voltage measurements of the LSMO/DDA samples and **(b)** 1D histograms at 2 V (313  I-V traces). **(c-d)** Same data after 2 h irradiation in blue 470 nm (112 I-V traces). **(e-f)** After 2 h irradiation in UV at 365*

*nm (2152 I-V traces).* ***(g-h)*** *After 2h blue irradiation at 470 nm (795 I-V traces).* ***(i)*** *Evolution of the log-mean current at +2 under the successive stimulations. A ratio of ca. 5 is observed. All C-AFM measurements under a flux of $N_2$ and loading force of 30 nN. Note that in this peculiar case, due to the high resistivity of the LSMO substrate, we did not measure current below 1 V and only positive bias were applied for this series of measurements.*

Due to a low quality of the SAMs for these two samples (ellipsometry thickness ~1nm compared to 1.7-2 nm for samples shown in Fig. 5, main text), the open/closed ratio are weak (<5) and we do not try to resolve two conductance peaks. It is likely that the worse quality for these two samples is related to the fact that DDA SAMs were grafted on previously used and recycled LSMO substrates (samples sonicated in DI water 5 min, in ethanol VLSI-grade 5 min, dried under N2 stream, followed by UV-ozone cleaning during 30 min and again the DI water/ethanol cleaning) instead of freshly fabricated LSMO substrate for the other samples.

## Effect of long time UV irradiations.

Figure S9 shows the 2D current histograms before and after the 365 nm irradiation (6 h) measured by C-AFM (ambient condition). Before irradiation, we measured a broad current distribution, which can also be decomposed in two peaks: P1 log μ = -11.92 ($1.2 \times 10^{-12}$ A) , P2 log μ = -9.59 ($2.6 \times 10^{-10}$ A) at -045 V (and almost the same values at 0.45 V). However, after the 365 nm irradiation, we observed a narrow distribution with only the peak P1 (Figs. S9d-f). Note also that only a small number of I-V traces have been recorded (46 compared to 521), which means that large parts of the sample have a current below the sensitivity of our C-AFM (1 pA). This result is surprising since we expect an increase of the molecular conductance when the diarylethenes are in the closed from.[9-12] Then, we exposed the sample to a 470 nm during 6h (switching DDA-c to DDA-o) and we remeasured the current (Figs. S9g-i). No change was observed.

Thus, no reversible optical switching of DDA is detected in the LSMO/DDA samples in this case. A possible expansion would be that long times (6h) irradiation at 365 nm can induce the blocked form of the diarylethne[13] for which we suppose a very low conductance.

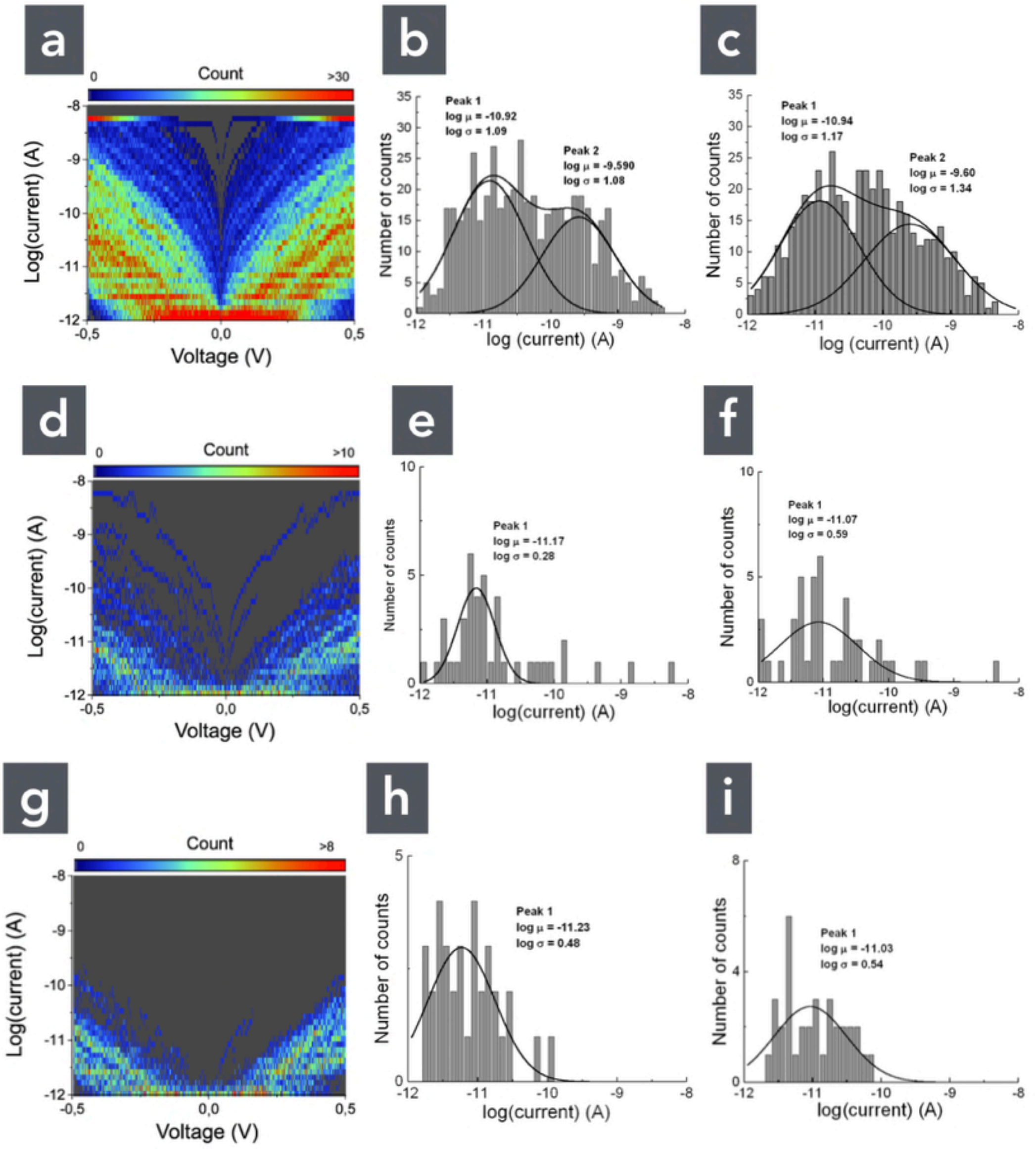

***Figure S9.*** ***(a)*** *2D histogram of current-voltage measurements of the LSMO/DDA sample (521 I-V traces) and* ***(b,c)*** *1D histograms at -0.45 V and +0.45 V.* ***(d)*** *2D histograms of current-voltage measurements of the DDA SAM after 6 h irradiation at 365 nm (46 I-V traces) and* ***(e,f)*** *1D histograms at at -0.45 V and +0.45 V.* ***(g)*** *2D histogram of current-voltage measurements of the LSMO/DDA sample after 6 h irradiation at 365 nm and 6h irradiation at 470 nm (34 I-V traces)*

*and **(h,i)** 1D histograms at -0.45 V and +0.45 V. All measurement by C-AFM in air and loading force of 30 nN. The black lines are the fits with a log-normal distribution. The fit parameters, log-mean current (log μ) and log-standard deviation (log σ) are given in the figures. Note that the relatively small number of I-V traces in these histograms (while around thousands were acquired, see details in the supporting information) is due to the fact that a majority of the traces showed currents <2x10$^{-12}$ A, close to the sensitivity and noise level of our apparatus, and were removed for clarity. Nevertheless, a number of counts >20-30 is reasonably sufficient to a significant statistical analysis of C-AFM measurements on SAMs.[14]*

## Effect of long time UV irradiations on a bare LSMO sample.

We irradiated a bare LSMO sample in the same conditions as in Fig. S9 (365 nm, 6h) and measured the I-V curves (Fig. S10). We observed a current decrease by a factor 16 from log μ ≈ -8.8 (1.6x10$^{-9}$ A) to -10 (10$^{-10}$ A). This persistent photoresistivity (PPR) effect in LSMO[15] - increase of LSMO resistance (a factor about 15) after a long time UV light irradiation - can also contribute to the total current decrease observed (Fig. S9) for the LSMO/DDA sample. We discard the possibility that a long time UV irradiation had severely degraded (removed) the SAM, because in that case, we would have observed, in the histograms Figs. S9d-f, a second current peak (with a significant number of counts) centered at around 10$^{-10}$ A as shown in Fig. S10d-f for the LSMO substrate alone. Albeit UV light can generate ozone, which is well known and frequently used to remove organic compounds on a surface, we note that we have used an "ozone-free" UV source (centered at 365 nm with a bandwidth of 10 nm, no UV light at 185 nm), which minimize the risk of SAM degradation.

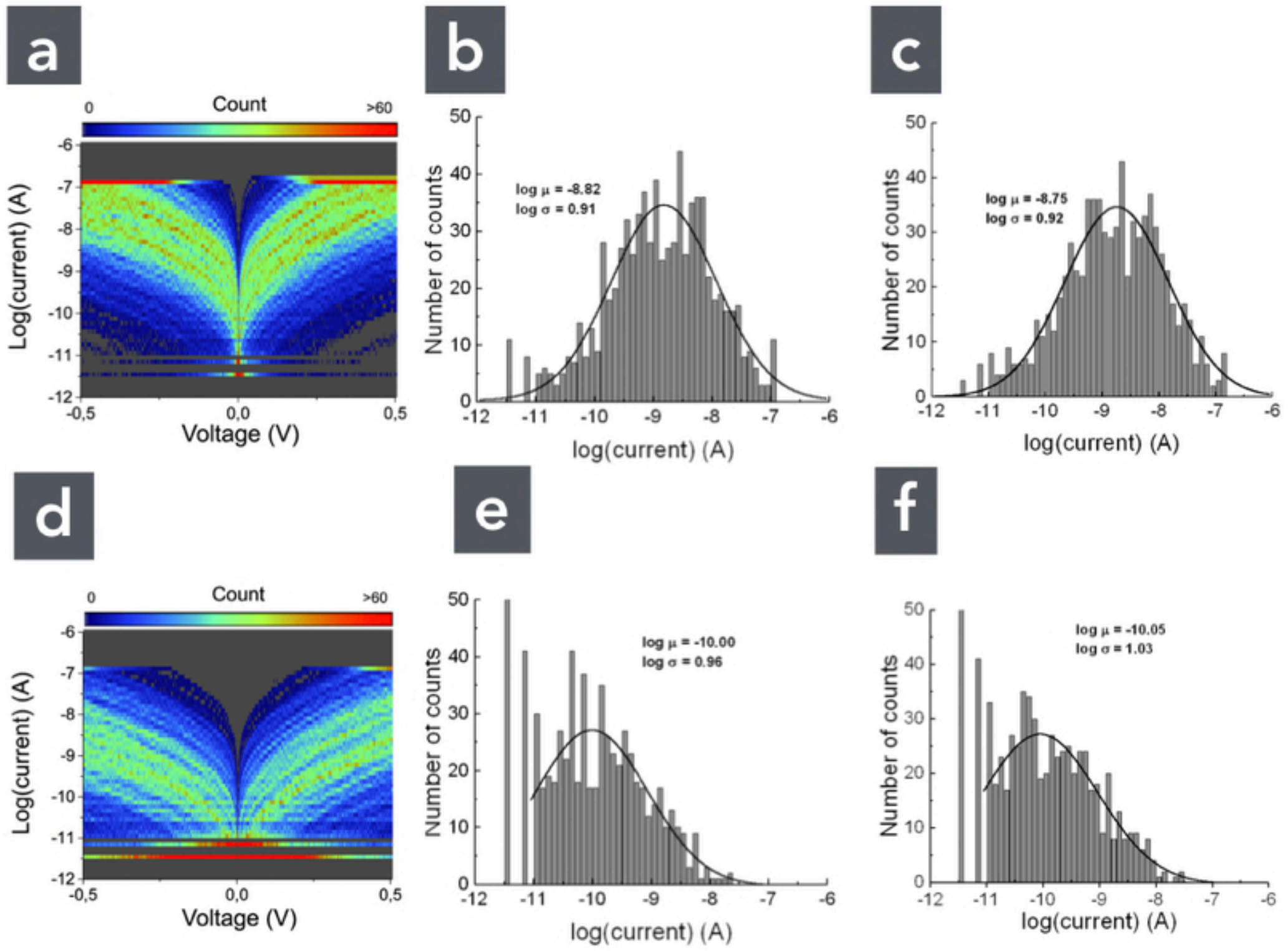

***Figure S10.*** ***(a-c)*** *2D histogram of current-voltage measurements of the LSMO substrate (800 I-V traces) and 1D histograms at -0.1 V and +0.1 V.* ***(d-f)*** *2D histogram of current-voltage measurements of the LSMO substrate after 6 h irradiation at 365 nm (800 I-V traces) and 1D histograms at -0.1 V and +0.1 V. All C-AFM measurements in air and loading force of 30 nN.*